\documentclass[aps,amsmath,superscriptaddress,amssymb,twocolumn,longbibliography,floatfix]{revtex4-2}

\usepackage{mathrsfs} 
\usepackage{graphicx,color} 
\usepackage{wasysym}
\usepackage{mathtools}
\usepackage{bigints}
\usepackage{bm}
\usepackage{threeparttable}
\usepackage[normalem]{ulem}
\usepackage{physics}
\usepackage[colorlinks=true]{hyperref}
\hypersetup{
  colorlinks=true,
  linkcolor=blue,
  filecolor=magenta,
  citecolor=blue,
  urlcolor=blue,
}

\newcommand{\ovl}{\overline}

\newcommand{\mcat}{m_{\rm cat}}
\newcommand{\mnut}{m_{\rm nut}}
\newcommand{\alphacat}{\alpha_{\rm cat}}
\newcommand{\alphanut}{\alpha_{\rm nut}}
\newcommand{\pin}{p_u^{\rm (in)}}
\newcommand{\pout}{p_v^{\rm (out)}}

\newcommand{\mP}[1]{\mathcal{P}\qty[#1]}
\newcommand{\reaction}{j+k_1+\cdots +k_\nu \to i+k_1+\cdots +k_\nu}
\newcommand{\cijk}{C_{ij}^{k_1 \ldots k_\nu}}
\newcommand{\cjik}{C_{ji}^{k_1 \ldots k_\nu}}

\begin{document}

\title{Multicatalyst reactions induce abrupt transition in a dense catalytic reaction network model}

\author{Kota Mitsumoto}
\email{kmitsumoto@g.ecc.u-tokyo.ac.jp}
\affiliation{Graduate School of Arts and Sciences, The University of Tokyo, Komaba 3-8-1, Meguro-ku, Tokyo 153-8902, Japan}
\author{Shuji Ishihara}%
\email{csishihara@g.ecc.u-tokyo.ac.jp}
\affiliation{Graduate School of Arts and Sciences, The University of Tokyo, Komaba 3-8-1, Meguro-ku, Tokyo 153-8902, Japan}
\affiliation{Universal Biology Institute, The University of Tokyo, Komaba 3-8-1, Meguro-ku, Tokyo 153-8902, Japan}

\date{\today}

\begin{abstract}
We generalize the Furusawa--Kaneko model, a simple model of intracellular catalytic reaction networks, to multicatalyst reactions and analyze the generalized model using dynamical mean-field theory. In the thermodynamic and dense-network limit, we derive exact effective equations and analyze their fixed points for arbitrary degree distributions. Under nutrient-poor conditions, network heterogeneity suppresses the metabolic--starvation transition irrespective of the number of catalysts per reaction.  In contrast, under nutrient-rich conditions, the bifurcation structure changes with the number of catalysts. With one catalyst per reaction, only continuous transitions occur, whereas two or more catalysts give rise to bistable phases and discontinuous transitions. The bifurcation structure also differs between two and three catalysts per reaction, with three catalysts allowing bistability between metabolic states with different cell growth rates. Our results provide a possible mechanism for bistability and abrupt switching in metabolic dynamics.
\end{abstract}

\maketitle

\section{Introduction}

Network representation is a powerful theoretical approach for modeling and understanding the dynamics of complex systems~\cite{albert2002statistical, barabasi2016network, newman2010networks, cohen2010complex}, including social~\cite{vespignani2011modelling, castellano2009statistical}, ecological~\cite{montoya2006ecological}, information~\cite{caldarelli2026physics}, neuronal~\cite{seguin2023brain}, and metabolic systems~\cite{barabasi2004network, ghosh2024exploring}. A network represents dynamical variables as nodes and their interactions as edges, allowing structural properties such as mean connectivity, degree distributions, and modularity to be characterized. How these structural properties affect the dynamics of such complex systems has been studied~\cite{boccaletti2006complex, dorogovtsev2008critical, barzel2013universality}. Beyond these conventional structural properties, attention has increasingly focused on hypergraph structures, in which three or more nodes are connected by a single hyperedge to represent higher-order interactions, and to their effects on dynamics~\cite{battiston2020networks, battiston2021physics}. In a variety of systems, phase transitions that are continuous with only pairwise interactions can become discontinuous when higher-order interactions are introduced, accompanied by bistability and hysteresis~\cite{gardner1985spin, sourlas1989spin-glass, wang1991chaos, kabashima2000typical, franz2001ferromagnet, davatolhagh2010nature, skardal2020higher, dearruda2020social, sun2023dynamic, anwar2024self-organized, dutta2024transition, cavaliere2026graphical}. These findings motivate the investigation of higher-order interactions as a possible mechanism underlying abrupt switching behavior in biological systems.


Metabolic networks are representative biological systems with higher-order interactions~\cite{klamt2009hypergraphs, marehallisrinivas2025hypergraph-based}. Chemical reactions generally involve the transformation of multiple reactant species into products, and in metabolic systems, they are typically catalyzed by another chemical species, so that multiple species participate in a single reaction. The Furusawa--Kaneko model~\cite{furusawa2003zipf, furusawa2006evolutionary, furusawa2012adaptation, kaneko2006life, kaneko2025universal} is a simplified metabolic network model that incorporates only the catalytic aspect of such reactions. Specifically, intracellular chemical reactions are represented as simple catalytic reactions $j+k\to i+k$, in which the chemical species $j$ is converted into $i$, catalyzed by $k$. Even under this simplification, each reaction still connects the three nodes $i$, $j$, and $k$ through a hyperedge, so that the model retains a hypergraph structure. This model also incorporates nutrient uptake through the membrane and dilution due to cell growth. Recently, the Furusawa--Kaneko model was solved exactly in the dense-connectivity and thermodynamic limits using dynamical mean-field theory (DMFT)~\cite{mitsumoto2026starvation}. DMFT is based on a generating-functional formalism for dynamical paths~\cite{martin1973statistical, janssen1976on, dedominicis1978dynamics, sompolinsky1982relaxational, cugliandolo1993analytical, castellani2005spin, galla2024generating} and enables ensemble averaging over random network topologies with arbitrary degree distributions~\cite{mimura2009parallel, poley2025interaction, metz2025dynamical}. The exact analysis in Ref.~\cite{mitsumoto2026starvation} revealed three distinct phases: a metabolic phase with sustained cell growth driven by catalytic reactions, an overnutrition phase in which intracellular composition consists solely of nutrients, and a starvation phase in which nutrient supply is insufficient to sustain cell growth. Comparison across different degree distributions showed that network heterogeneity promotes cell growth under nutrient-poor conditions, thereby suppressing starvation. In particular, for scale-free topology, a common structural feature of metabolic networks characterized by highly heterogeneous power-law degree distributions~\cite{jeong2000the, jeong2001lethality, albert2002statistical, barabasi2004network, barabasi2016network, broido2019scale}, the starvation phase disappears from the phase diagram. By contrast, the metabolic--overnutrition transition at high nutrient supply is insensitive to the degree distribution, with its phase boundary remaining unchanged across different degree distributions. Moreover, it has also been shown that, despite the presence of higher-order interactions, all phase transitions are continuous irrespective of the degree distribution.

\begin{figure}[t]
\centering
\includegraphics[width=80mm]{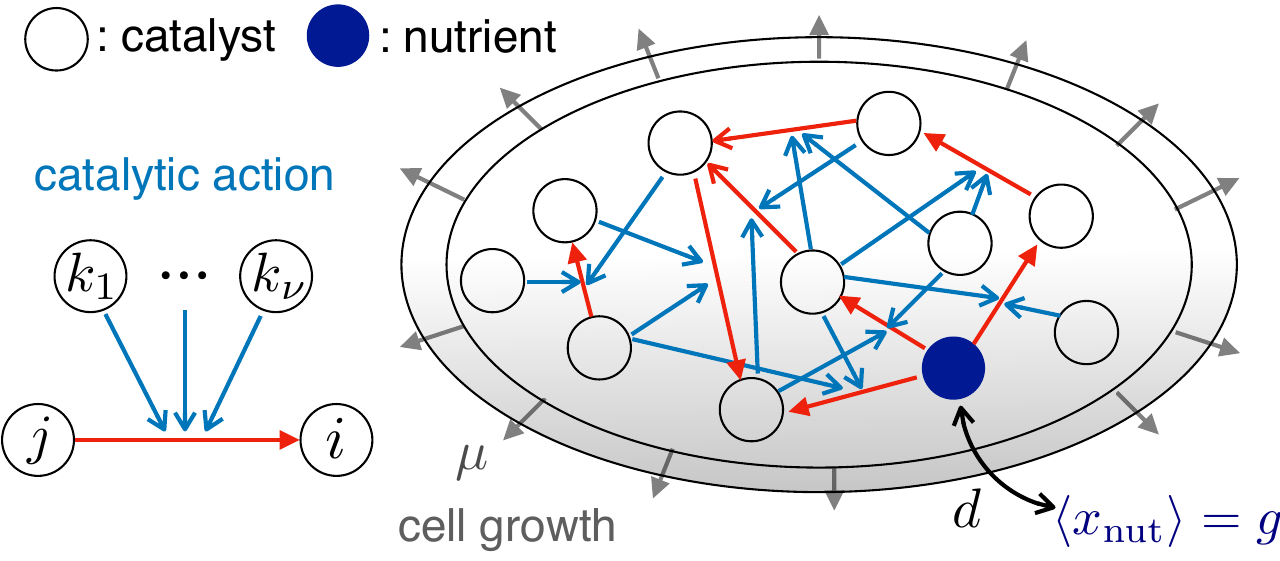}
\caption{
Schematic illustration of a model of a cell incorporating multicatalyst reactions, nutrient uptake, and cell growth. The catalytic reaction $\reaction$ represents the conversion of the chemical species $j$ into the $i$ catalyzed by the $k_1,\ldots,k_\nu$. The $\nu=2$ case is shown as an example. The model consists of catalytic chemical species and nutrients. Nutrients are catalytically inactive and can pass through the cell membrane with the permeability $d$. The equilibrium abundance of nutrients in the external environment is $g$. The cell grows at a rate $\mu$, which is also a dynamical variable. 
}
\label{fig:schematic}
\end{figure}

In this paper, we analyze the Furusawa--Kaneko model with multicatalyst reactions (see Fig.~\ref{fig:schematic}) using DMFT, and show that multicatalyst reactions give rise to bistable phases and discontinuous phase transitions. Metabolic networks often contain sequential reaction pathways in which different enzymes catalyze successive reaction steps~\cite{ramirezgaona2016YMDB, haug2020metaboLights, wishart2022HMDB, kim2025PubChem, kanehisa2025KEGG}. These enzymes have also been suggested to form multienzyme complexes, bringing the enzymes acting on successive reactions into close proximity~\cite{srere1987complexes, kohnhorst2017identification, pareek2021metabolic}. As a coarse-grained representation of such reactions, we generalize the catalytic reaction in the Furusawa--Kaneko model to $\reaction$, in which the conversion of $j$ into $i$ is catalyzed by $\nu$ chemical species. The case $\nu=1$ corresponds to the original Furusawa--Kaneko model. The case $\nu=2$ is illustrated schematically in Fig.~\ref{fig:schematic}. Using DMFT for arbitrary degree distributions, we derive effective dynamical equations conditioned on the in- and out-degrees, which become exact in the dense-connectivity and thermodynamic limits. As in Ref.~\cite{mitsumoto2026starvation}, we analyze the fixed-point structure of networks with Poisson, uniform, and power-law degree distributions. Under nutrient-poor conditions, starvation is suppressed as the network becomes more heterogeneous, irrespective of $\nu$, and the starvation phase disappears for the power-law out-degree distribution. The metabolic--starvation transition observed for the Poisson and uniform distributions is continuous for all $\nu$. Under nutrient-rich conditions, by contrast, the phase behavior changes qualitatively with $\nu$. For $\nu\ge2$, a bistable phase appears in which the metabolic and overnutrition states coexist. The transition from the metabolic state to the overnutrition state is discontinuous, and once the system falls into the overnutrition state, reversing the parameter change does not restore the metabolic state. For $\nu\ge3$, another bistable phase emerges in which two metabolic states with different cell growth rates coexist, giving rise to discontinuous switching between distinct growth states as the parameters are varied. 
This qualitative phase behavior is common to the Poisson, uniform, and power-law degree distributions. We also perform numerical simulations at finite mean connectivity (see Appendix Sec.~\ref{sec:simulation_methods} for numerical methods) and confirm that the mean-field theory provides a good description of finite-connectivity systems.

This paper is organized as follows. In Sec.~\ref{sec:model}, we describe the model in detail. In Sec.~\ref{sec:effective}, we present the effective mean-field equations and their fixed-point solutions, with the detailed derivation given in Appendix Sec.~\ref{sec:dmft_derivation}. In Sec.~\ref{sec:bifurcation}, we provide a comprehensive analysis of the bifurcation structure for the Poisson degree distribution at arbitrary $\nu$. In Sec.~\ref{sec:phase}, we present phase diagrams for the Poisson, uniform, and power-law degree distributions for $\nu=1$, $2$, and $3$. Sec.~\ref{sec:starvation} is devoted to the analysis of nutrient-poor conditions for $\nu=2$ and $3$, where we show that starvation is suppressed as the degree distribution becomes more heterogeneous and disappears for the power-law distribution. In Sec.~\ref{sec:bistability}, we investigate the physical quantities associated with the two bistable phases that emerge for $\nu=3$ under nutrient-rich conditions. Finally, Sec.~\ref{sec:summary} summarizes our results.


\section{Model}
\label{sec:model}
We consider a generalized Furusawa--Kaneko model, in which a cell is represented as a membrane-bound compartment containing $N$ chemical species that interact through multicatalyst reactions on a random network, as illustrated in Fig.~\ref{fig:schematic}. The abundance of each chemical species is denoted by $x_i(t)$ for $i=1,2,...,N$. The multicatalyst reactions in this model take the simple form $\reaction$, representing the conversion of chemical species $j$ into $i$, catalyzed by $k_1,\ldots,k_\nu$. In this reaction, the production rate of $x_i$ and consumption rate of $x_j$ are both given by $x_j x_{k_1}\cdots x_{k_\nu}$, so that the total abundance $x_i + x_j$ is conserved. The chemical species are divided into two groups: catalytic chemical species and nutrients, denoted by $\mathcal{G}_{\rm cat}$ and $\mathcal{G}_{\rm nut}$, respectively. Their fractions are denoted by $\alphacat$ and $\alphanut$, satisfying $\alphacat + \alphanut = 1$. Nutrients are assumed to be catalytically inactive. In Ref.~\cite{furusawa2003zipf, furusawa2006evolutionary, mitsumoto2026starvation}, catalytic chemical species are further classified into membrane-penetrable and membrane-unpenetrable species, with the former able to diffuse out of the cell. However, previous studies have shown that this additional classification does not qualitatively change the behavior of the model~\cite{furusawa2006evolutionary, mitsumoto2026starvation}. We therefore assume here that all catalytic chemical species are unpenetrable.

Nutrients can diffuse across the membrane and exchange with the external environment at rate $d(g-x_i)$, respectively, where $d$ is the membrane permeability, and $g$ is the equilibrium abundance of nutrients in the external environment. For convenience, we introduce parameters $(d_i, g_i)$, defined as $(0,0)$ for $i \in \mathcal{G}_{\rm cat}$ and $(d,g)$ for $i \in \mathcal{G}_{\rm nut}$. Nutrient uptake drives cell growth, which uniformly dilutes the abundances of intracellular chemical species by a term $-\mu x_i$, where $\mu = \frac{1}{N}\sum_{i =1}^N d_i(g_i-x_i)$ is the cell growth rate. Here and below, the abundance $x_i$, time $t$, permeability $d$, equilibrium nutrient abundance $g$, and growth rate $\mu$ are dimensionless; we follow the nondimensionalization procedure in Ref. \cite{mitsumoto2026starvation}, which assumes that the cell-volume growth is determined by the net increase in the total amount of chemical species, which leads to the above definition of the cell growth rate (see Appendix Sec.~\ref{sec:nondimension}).

Taken together, the dynamics of $x_i(t)$ is governed by the following equations:
\begin{align}
\nonumber
\dot{x}_i(t) &= \frac{1}{c}\sum_{j =1}^N \sum_{\{k\} \in \mathcal{G}_{\rm cat}} \cijk x_j(t) x_{k_1}(t)\cdots x_{k_\nu}(t) \\ \nonumber
 &-  \frac{1}{c}\sum_{j =1}^N \sum_{\{k\} \in \mathcal{G}_{\rm cat}} \cjik x_i(t) x_{k_1}(t)\cdots x_{k_\nu}(t)  \\
 &+ d_i(g_i-x_i(t)) - \mu(t) x_i(t),
 \label{eq:dynamical1}
\end{align}
where $\sum_{\{k\} \in \mathcal{G}_{\rm cat}}$ denotes the summation over all tuples $(k_1,\ldots,k_\nu)$ such that each $k_i$ belongs to $\mathcal{G}_{\rm cat}$. The first and the second terms represent the catalytic reactions incoming to and outgoing from the $i$-th chemical species, respectively. The tensor $\{ \cijk \}$ encodes the topology of catalytic reaction network, where $C_{ij}^{k_1 \ldots k_\nu}$ is equal to $1$ if the  reaction $\reaction$ exists and $0$ otherwise. $c$ is the mean connectivity of the network, i.e., $\frac{1}{N}\sum_{i,j,k_1,\ldots, k_\nu}\cijk= c$. The factor $1/c$ normalizes the catalytic reaction terms so that the reaction flux into each species remains $\mathcal{O}(1)$ in the large-$c$ limit. This normalization allows networks with different values of $c$ to be compared on the same reaction-rate scale and to calculate the dense-connectivity limit exactly using DMFT (see Appendix Sec.~\ref{sec:dmft_derivation}). The third and fourth terms represent nutrient exchange with the environment and dilution due to cell growth, respectively. By summing Eq.~\eqref{eq:dynamical1} over $i$, one finds $\sum_{i=1}^N \dot{x}_i(t) = \mu(t) (N-\sum_{i=1}^N x_i(t))$; hence, positive $\mu(t)$ enforces a constraint on the long-time behavior of the average abundance of chemical species, leading to
\begin{align}
    \lim_{t\to \infty}\frac{1}{N}\sum_{i=1}^N x_i(t) = 1.
    \label{eq:constraint1}
\end{align}
If the growth rate becomes negative, the cell instead shrinks and eventually dies. In the model, we assume constant catalytic reaction rates following previous studies~\cite{furusawa2003zipf, furusawa2006evolutionary, mitsumoto2026starvation}. We also do not explicitly model cell division, since its inclusion does not affect the conclusions of the analysis below.


To derive the effective dynamical equations for random networks, we adopt the configuration model~\cite{bollobs1980a, chung2002connected, newman2010networks, poley2025interaction}, which allows us to construct random networks with arbitrary distributions of in-degrees $\pin$ and out-degrees $\pout$, both having mean $c$. Here, the in-degree and out-degree of a node count the numbers of incoming and outgoing reaction edges, respectively. To construct a network, we first independently draw the in-degree $u_i$ and out-degree $v_i$ for each node from distributions $\pin$ and $\pout$, respectively. For each combination $(i,j,k_1,\ldots,k_\nu)$, the connection $\cijk$ is then set to $1$ with probability $\frac{u_iv_j}{Nc(N\alphacat)^{\nu}}$ and to $0$ otherwise. The factor $1/(N\alphacat)^{\nu}$ accounts for the random selection of the $\nu$ catalysts from the $N\alphacat$ catalytic chemical species in $\mathcal{G}_{\rm cat}$. For sufficiently large $N$, this construction yields networks with the prescribed degree distributions. Below, we consider three kinds of distributions: a Poisson distribution, $p_x=c^x e^{-c}/x!$ (corresponding to the Erd\H{o}s--R\'{e}nyi (ER) network~\cite{erdos1959random}), a uniform distribution, $p_x = 1/2rc$ defined on the interval $[(1-r)c, (1+r)c]$ with $0<r\le 1$, and a power-law distribution, $p_x = 2c(c+1)/(x+c)(x+c+1)(x+c+2)\sim x^{-3}$~\cite{dorogovtsev2000structure}. Introducing the rescaled degree $y=x/c$, in the dense-network limit $c \gg 1$, the corresponding distributions converge to $\delta(y-1)$, $1/2r$ for $1-r\le y\le1+r$, and $2/(y+1)^3$ for $y\ge0$, respectively. Thus, networks with uniform or power-law degree distributions retain structural heterogeneity even in the dense limit, whereas the ER network approaches a regular (homogeneous) random network.


\section{Effective dynamical equations and fixed points}
\label{sec:effective}

In the thermodynamic limit, $N\to\infty$, and for dense connectivity, $c \gg 1$, we analyze the dynamics governed by Eq.~\eqref{eq:dynamical1} using DMFT. By averaging over the network topology for given degree distributions, we derive exact effective dynamical equations for the typical abundances of catalytic chemical species and nutrients with in-degree $u$ and out-degree $v$, denoted by $x_{{\rm cat},uv}(t)$ and $x_{{\rm nut},uv}(t)$, respectively. The resulting dynamics is independent of the catalytic degree because the catalyzed chemical species do not feed back into its own dynamics through the catalytic activity.  The derivation of the DMFT equations is presented in Appendix Sec.~\ref{sec:dmft_derivation}. The effective dynamical equations are given by
\begin{align}
\nonumber
\dot{x}_{{\rm cat},uv}(t) &= \frac{u}{c}m_{\rm in}(t) \mcat^\nu(t) - \frac{v}{c} \mcat^\nu(t) x_{{\rm cat},uv}(t) \\
&- \mu(t) x_{{\rm cat},uv}(t), 
\label{eq:dynamical_cat} \\ \nonumber
\dot{x}_{{\rm nut},uv}(t) &= \frac{u}{c}m_{\rm in}(t) \mcat^\nu(t) - \frac{v}{c} \mcat^\nu(t) x_{{\rm nut},uv}(t) \\
&+ d(g-x_{{\rm nut},uv}(t)) - \mu(t) x_{{\rm nut},uv}(t),
\label{eq:dynamical_nut}
\end{align}
where $m_{\rm in}(t) = \sum_{u,v} \frac{v}{c} \pin\pout  (\alphacat x_{{\rm cat},uv}(t)+\alphanut x_{{\rm nut},uv}(t))$. $m_{\rm cat}(t) = \sum_{u,v} \pin\pout x_{{\rm cat},uv}(t)$ and $m_{\rm nut}(t) = \sum_{u,v} \pin\pout x_{{\rm nut},uv}(t)$ represent the average abundances of catalytic chemical species and nutrients, respectively. The growth rate is given by $\mu(t) = d \alphanut (g  - \mnut(t))$. By averaging Eqs.~\eqref{eq:dynamical_cat} and \eqref{eq:dynamical_nut} over $u$ and $v$, and using the definition of $m_{\rm in}(t)$, one finds that the average abundance of all chemical species converges to
\begin{align}
\lim_{t \to \infty}(\alphacat \mcat(t) + \alphanut \mnut(t)) = 1
\label{eq:constraint2}
\end{align}
if $\mu(t)>0$, corresponding to Eq.~\eqref{eq:constraint1}. When Eq. \eqref{eq:constraint2} is satisfied, the growth rate $\mu$ and the average abundance of nutrient $\mnut$ can be expressed in terms of $\mcat$ as
\begin{align}
\mu(t) = d(\alphanut g - 1 + \alphacat \mcat(t)), \label{eq:growth1} \\
\mnut(t) = (1-\alphacat \mcat(t))/\alphanut, \label{eq:mnut1}
\end{align}
respectively. 


At the fixed point, the following relations hold:
\begin{align}
    x_{{\rm cat}, uv}^* &= \frac{\frac{u}{c} m_{\rm in}^* (\mcat^*)^\nu}{\mu^* + \frac{v}{c} (\mcat^*)^\nu}, \label{eq:fixed_cat} \\
    x_{{\rm nut}, uv}^* &= \frac{d g + \frac{u}{c} m_{\rm in}^* (\mcat^*)^\nu}{\mu^* + d + \frac{v}{c} (\mcat^*)^\nu}, \label{eq:fixed_nut}
\end{align}
where $A^* = \lim_{t\to\infty}A(t)$. These expressions show that the chemical abundances increase with increasing in-degree $u$ and decreases with increasing out-degree $v$. By averaging $x_{{\rm cat}, uv}^*$ and $x_{{\rm nut}, uv}^*$ over $u$ and $v$, we obtain 
\begin{align}
\mcat^* &= m_{\rm in}^* (\mcat^*)^\nu X(\mu^*, (\mcat^*)^\nu), \label{eq:mcat1} \\
\mnut^* &= (dg + m_{\rm in}^* (\mcat^*)^\nu)X(\mu^*+d, (\mcat^*)^\nu), \label{eq:mnut2}
\end{align}
where $X(a,b) = \sum_{v}\pout/(a+\frac{v}{c}b)$. At $\nu=1$, these equations coincide with those derived in Ref.~\cite{mitsumoto2026starvation}. Notably, $\mcat^*$ and $\mnut^*$ do not depend on the in-degree distribution; hence, macroscopic quantities such as $\mu^*$ are independent of the in-degree distribution. Substituting Eqs.~\eqref{eq:growth1} and \eqref{eq:mnut1} into Eqs.~\eqref{eq:mcat1} and \eqref{eq:mnut2}, we obtain a closed set of equations for $\mcat^*$ and $m_{\rm in}^*$. We numerically solve these equations for the Poisson, uniform, and power-law out-degree distributions.


\begin{figure*}[t]
\centering
\includegraphics[width=170mm]{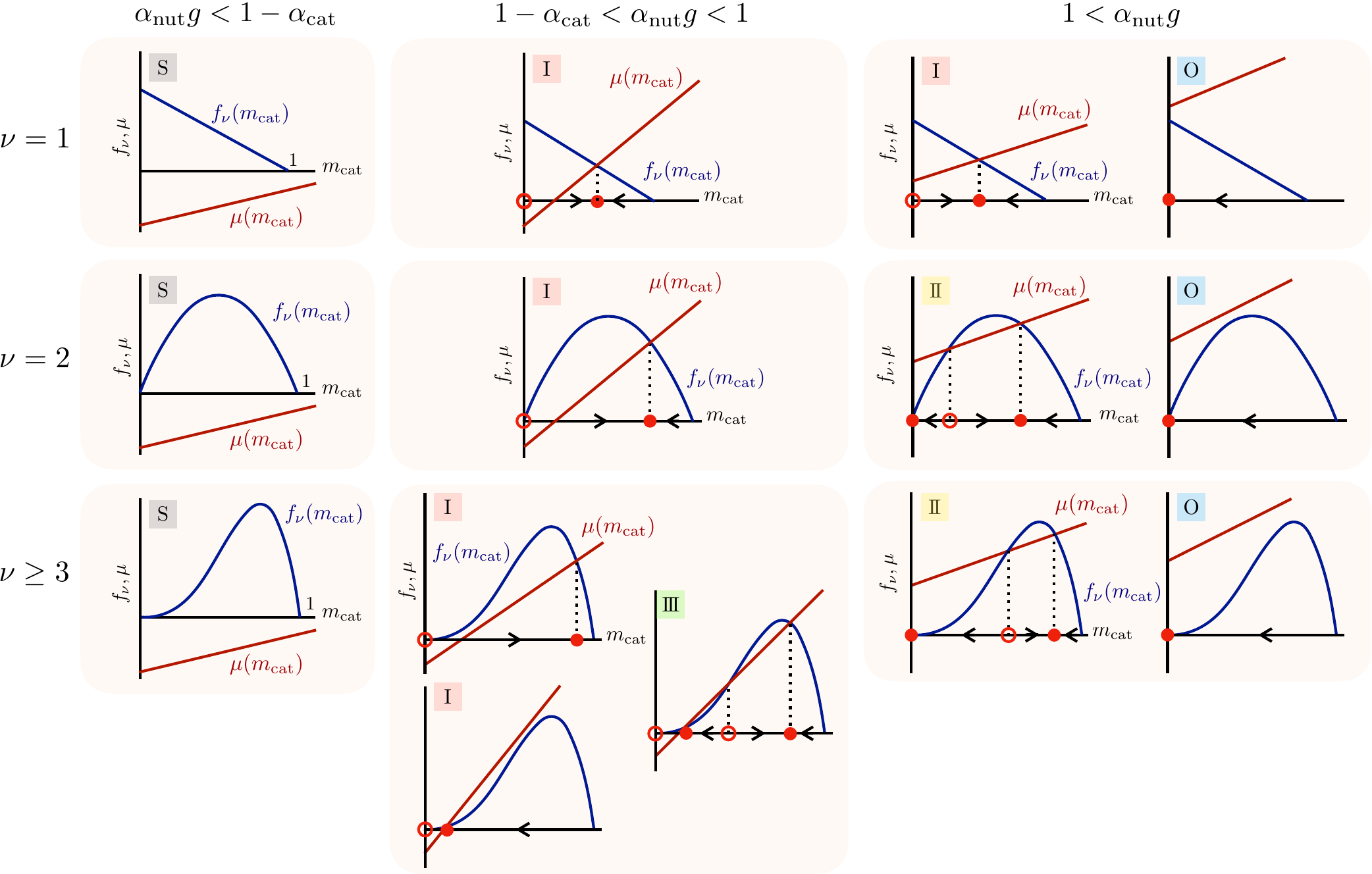}
\caption{
Bifurcation diagram of $\mcat$ in the dense-network limit for a Poisson out-degree distribution when $\alphacat \mcat(t)+\alphanut \mnut (t) = 1$. The solid and open circles on the horizontal axis represent stable and unstable fixed points $\mcat^*$. The black arrows indicate the flow of $\mcat$, which is determined by whether $\mu(\mcat) = d(\alphanut g -1 +\alphacat \mcat)$ is larger or smaller than $f_{\nu}(\mcat)=\mcat^{\nu-1}(1-\mcat)$. The phases S, O, I, II, and III correspond to the starvation phase, overnutrition phase, metabolic phase, bistable phase between metabolic and overnutrition states, and bistable phase between high-growth metabolic and low-growth metabolic states, respectively.
}
\label{fig:schematic_dyna}
\end{figure*}

\section{Bifurcation structure for the Poisson distribution}
\label{sec:bifurcation}

Before examining the dependence on the out-degree distribution, we first analyze the bifurcation structure for the Poisson distribution. In the dense-network limit, the Poisson out-degree distribution approaches $\delta(v/c-1)$, and $m_{\rm in}(t)$ coincides with the average abundance of all chemical species. If the initial condition is chosen such that the average abundance of all chemical species is one, $m_{\rm in}(t)=1$ holds at all times as long as $\mu(t)>0$. By averaging Eq.~\eqref{eq:dynamical_cat} over $u$ and $v$ and using Eq.~\eqref{eq:growth1}, we obtain
\begin{align}
\dot{m}_{\rm cat}(t) &= \mcat(t) \qty[\mcat^{\nu-1}(t)-\mcat^{\nu}(t)-\mu(\mcat(t))] \nonumber \\
&=\mcat(t) \qty[f_{\nu}(\mcat(t))-\mu(\mcat(t))],
\end{align}
where we define $f_{\nu}(\mcat) = \mcat^{\nu-1}(1-\mcat)$. The fixed points are given by $\mcat^*=0$ and the solutions of $\mu(\mcat^*) = f_{\nu}(\mcat^*)$, and the direction of the flow is determined by the sign of $f_{\nu}(\mcat)-\mu(\mcat)$. We refer to the states corresponding to $\mcat^*=0$ and $\mu(\mcat^*) = f_{\nu}(\mcat^*)$ as the overnutrition and metabolic states, respectively. The growth rate in the metabolic state is given by $\mu(\mcat^*) = f_{\nu}(\mcat^*)$, whereas that in the overnutrition state is $\mu^* = d(\alphanut g-1)$. 

\begin{figure*}[t]
\centering
\includegraphics[width=170mm]{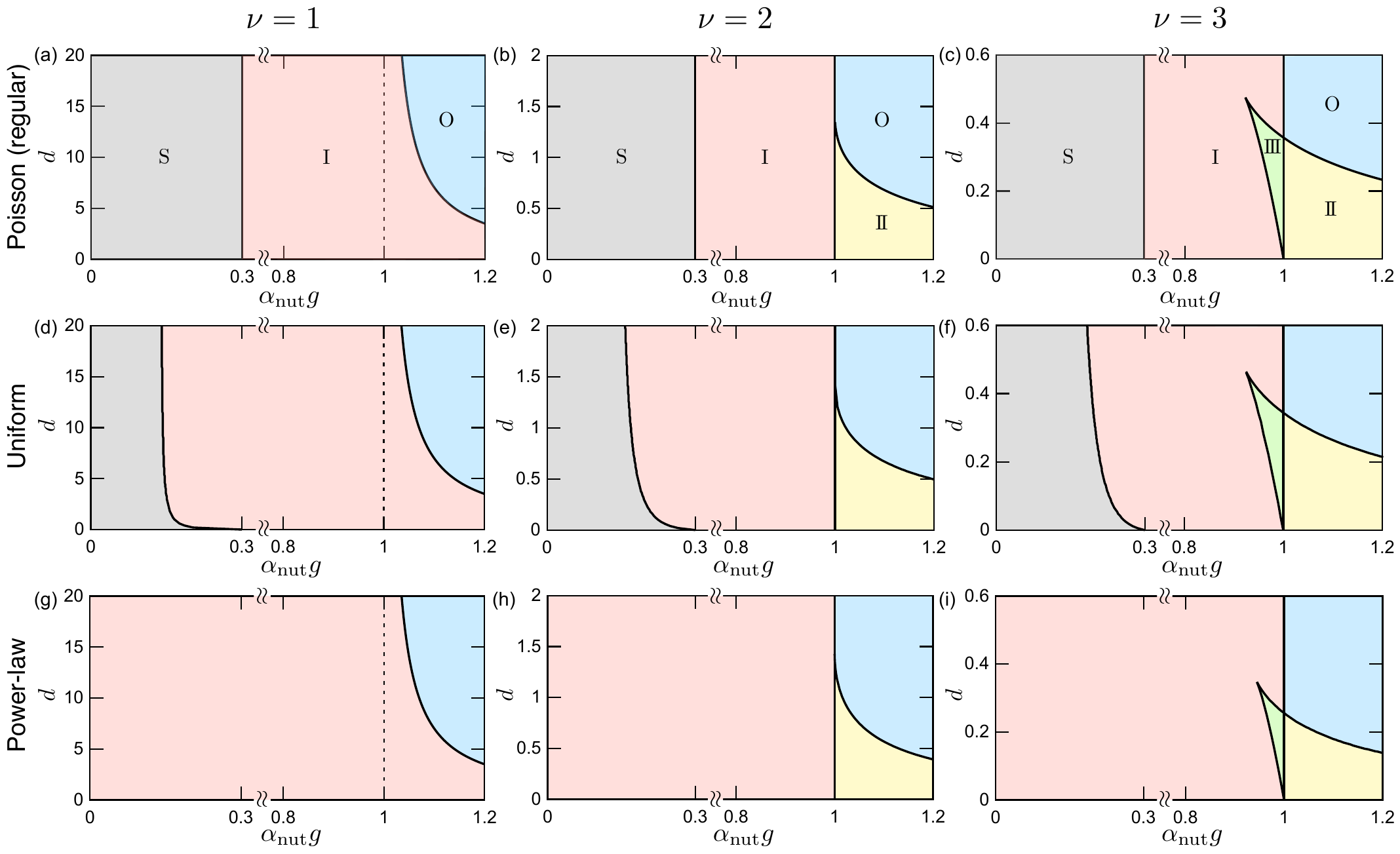}
\caption{
Phase diagrams in the dense-network limit for (a) $\nu=1$, (b) $\nu=2$, and (c) $\nu=3$ with a Poisson out-degree distribution; (d) $\nu=1$, (e) $\nu=2$, and (f) $\nu=3$ with a uniform out-degree distribution ($r=0.99$); (g) $\nu=1$, (h) $\nu=2$, and (i) $\nu=3$ with a power-law out-degree distribution. The phases S, O, I, II, and III correspond to the starvation phase, overnutrition phase, metabolic phase, bistable phase between metabolic and overnutrition states, and bistable phase between high-growth metabolic and low-growth metabolic states, respectively. The parameter $\alpha_{\rm cat} = 0.7$ is used.
}
\label{fig:phase}
\end{figure*}

When $\mu<0$, the relation in Eq.~\eqref{eq:constraint2} no longer holds; the cell shrinks and eventually dies. We therefore seek fixed-point solutions with $\mu>0$. The growth rate of the overnutrition state is positive only when $\alphanut g>1$. Thus, for $\alphanut g<1$, a positive-growth fixed point, if it exists, must be a metabolic state. However, as shown in the left column of Fig.~\ref{fig:schematic_dyna}, when $\alphanut g<1-\alphacat$, $\mu(\mcat)$ and $f_{\nu}(\mcat)$ have no intersection with $\mu(\mcat)>0$; hence, no positive-growth fixed-point solution exists in this parameter region. We refer to this region as the starvation phase, or Phase~S (see also Figs.~\ref{fig:phase}(a)--(c) for the phase diagrams).

For $\alphanut g>1$, the overnutrition state has a positive growth rate, and a bifurcation between the overnutrition and metabolic states can occur. For $\nu=1$, as reported in Ref.~\cite{mitsumoto2026starvation}, a transcritical bifurcation occurs between the two states. As illustrated in the top-right panel in Fig.~\ref{fig:schematic_dyna}, for $d>d_{\rm c}$, where $d_{\rm c}=1/(\alphanut g-1)$, $\mu(\mcat)$ and $f_1(\mcat)=1-\mcat$ have no intersection at $\mcat>0$, and the overnutrition state is stable. As $d$ decreases below $d_{\rm c}$, the fixed point at $\mcat=0$ becomes unstable, while the stable fixed point shifts continuously to $\mcat>0$, corresponding to the metabolic state. We refer to the parameter regions where the overnutrition state is uniquely stable and where the metabolic state is uniquely stable as Phase O and Phase I, respectively. For $\nu\geq2$, the bifurcation structure is qualitatively different, as shown in the middle and bottom panels of the right column in Fig.~\ref{fig:schematic_dyna}. For sufficiently large $d$, $\mu(\mcat)$ and $f_\nu(\mcat)$ have no intersection at $\mcat>0$, and only the overnutrition state is stable, corresponding to Phase O. As $d$ decreases, a pair of intersections emerges at finite $\mcat$. The fixed point at $\mcat=0$ and the intersection at larger $\mcat$ are stable, while the intersection at smaller $\mcat$ is unstable, leading to bistable phase distinct from Phase I. We refer to this bistable phase, in which the overnutrition and metabolic states coexist, as Phase~II (see also Figs.~\ref{fig:phase}(a)--(c)).

For $1-\alphacat<\alphanut g<1$, no bifurcation between the overnutrition and metabolic states occurs. For $\nu=1$ and $2$, as illustrated in the top and middle panels of the middle column in Fig.~\ref{fig:schematic_dyna}, $\mu(\mcat)$ and $f_\nu(\mcat)$ have only one intersection for any $d$, and thus only a single stable metabolic state exists in this parameter region, corresponding to Phase I. For $\nu\geq3$, however, bifurcations between different metabolic states can occur when $\alphanut g$ is close to one, as shown in the bottom-middle panel in Fig.~\ref{fig:schematic_dyna}. For sufficiently large or small $d$, $\mu(\mcat)$ and $f_\nu(\mcat)$ have only one intersection, whereas for intermediate $d$, three intersections appear. The intersections at the smallest and largest $\mcat$ are stable, while the intermediate one is unstable. Consequently, two metabolic states with different growth rates coexist as stable fixed points: the state with smaller $\mcat$ has a lower growth rate, while that with larger $\mcat$ has a higher growth rate. We refer to this bistable region as Phase~III (see also Figs.~\ref{fig:phase}(a)--(c)).

For the Poisson degree distribution, the bifurcation structure can be understood from the shape of $f_\nu(\mcat)$ in the interval $0 \leq \mcat \leq1$. The qualitative difference between $\nu=1$ and $\nu\geq2$ originates from whether $f_\nu(\mcat)$ is monotonically decreasing in this interval. The difference between $\nu=2$ and $\nu\geq3$, on the other hand, arises from the absence or presence of an inflection point. For any $\nu\geq3$, $f_\nu(\mcat)$ has a single inflection point in the interval $0<\mcat<1$; hence, increasing $\nu$ beyond $3$ does not introduce any qualitatively new bifurcation structure.


\section{Phase diagram}
\label{sec:phase}

In the previous section, we analyzed the bifurcation structure for the Poisson out-degree distribution in detail. For other out-degree distributions, the fixed points can be obtained by solving Eqs.~\eqref{eq:mcat1} and \eqref{eq:mnut2}. One immediately finds from Eq.~\eqref{eq:mcat1} a solution with $\mcat^* = 0$, corresponding to the overnutrition state. For this solution, $x_{{\rm cat}, uv}^*$ vanishes, while $x_{{\rm nut}, uv}^*=1/\alphanut$ for all $u$ and $v$; hence, the overnutrition state is independent of the out-degree distribution. In contrast, the metabolic state depends on the out-degree distribution; consequently, the phase diagram also varies with the out-degree distribution, as shown below.

Figure~\ref{fig:phase} summarizes the phase diagrams for the Poisson, uniform, and power-law out-degree distributions for $\nu=1$, 2, and 3. The phase behavior strongly depends on the out-degree distribution at low $\alphanut g$, i.e., nutrient-poor conditions. As shown in Figs.~\ref{fig:phase}(a)--(c), for the Poisson distribution, Phase S emerges when $\alphanut g<1-\alphacat$, irrespective of $\nu$. As the network becomes heterogeneous, Phase S is suppressed. For the uniform distribution, the Phase S region is smaller than that for the Poisson distribution, as shown in Figs.~\ref{fig:phase}(d)--(f), and for the power-law distribution, Phase S disappears for any $\nu$. Therefore, the suppression of starvation by network heterogeneity is robust, irrespective of $\nu$. In Sec.~\ref{sec:starvation}, we examine the behavior of physical quantities under nutrient-poor conditions and the mechanism underlying this suppression.

In contrast to the nutrient-poor regime, the phase behavior under nutrient-rich conditions exhibits only quantitative changes with the out-degree distribution (see Fig.~\ref{fig:phase}). For $\nu=1$, as reported in Ref.~\cite{mitsumoto2026starvation}, the phase boundary between Phases O and I is completely independent of the out-degree distribution. This is because the boundary is determined by the instability of the overnutrition state. For $\nu\geq2$, however, the overnutrition state remains stable throughout the region $\alphanut g>1$ for any out-degree distribution (see Appendix Sec.~\ref{sec:stability}), and hence Phase~II appears commonly. The boundary between Phases O and II for $\nu\geq2$ is determined by the instability of the metabolic state and therefore slightly depends on the out-degree distribution. For $\nu=3$, the phase boundary between Phases I and III emerges as a continuation of the boundary between Phases O and II. Although the qualitative structure of the phase diagram is unchanged by the out-degree distribution, Phase~III narrows as the network becomes more heterogeneous. The detailed behavior of the physical quantities around these bistable phases is examined in Sec.~\ref{sec:bistability}.


\begin{figure}[t]
\centering
\includegraphics[width=85mm]{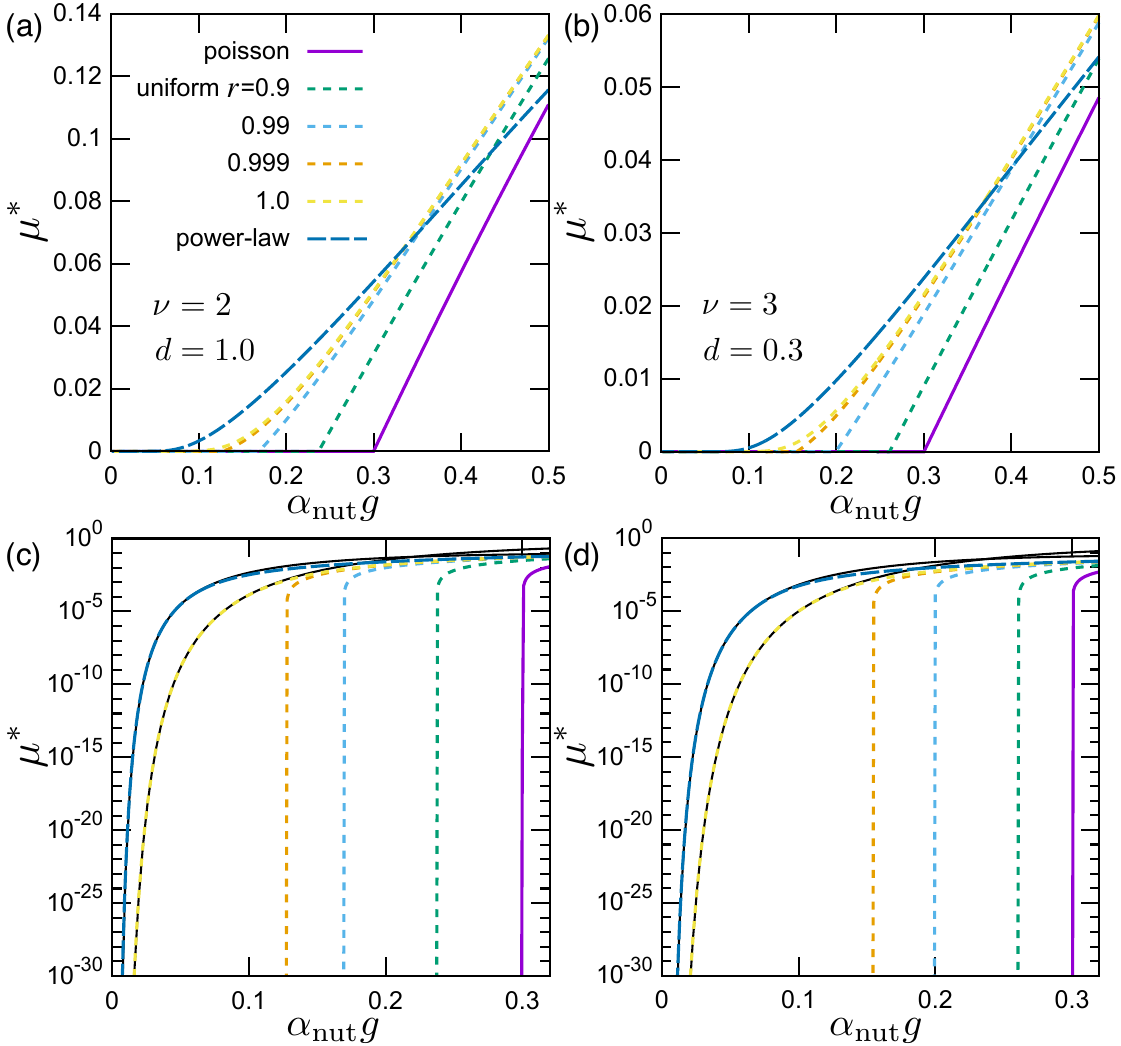}
\caption{
The growth rate $\mu^*$ against $\alpha_{\rm nut}g$ under nutrient-poor conditions, obtained from the DMFT for (a) $\nu=2$ and (b) $\nu=3$, and the semi-log plots for (c) $\nu=2$ and (d) $\nu=3$.  Different curves represent different network topologies: results for Poisson, uniform ($r=0.9,0.99,0.999,1$), and power-law out-degree distributions are shown. The black solid curves in panels (c) and (d) represent the asymptotic solutions for $\alpha_{\rm nut}g \ll 1$. The parameter $\alphacat = 0.7$ is commonly used and $d=1.0$ and $0.3$ are used for $\nu=2$ and $\nu=3$, respectively.
}
\label{fig:starvation}
\end{figure}

\section{Starvation suppression}
\label{sec:starvation}

We now examine the DMFT results under low nutrient supply. For $\nu=1$, it has been reported in Ref.~\cite{mitsumoto2026starvation} that more heterogeneous networks exhibit higher cell growth rates. In particular, for the uniform distribution with $r=1$ and the power-law distribution, the growth rate remains positive for any $\alphanut g$, leading to the absence of the metabolic--starvation transition. As shown in Sec.~\ref{sec:phase}, the starvation phase also disappears for the power-law out-degree distribution for $\nu=2$ and $3$. Here, we examine how the cell growth rate behaves under nutrient-poor conditions for $\nu=2$ and $3$.

Figures~\ref{fig:starvation} (a) and (b) show the dependence of the growth rate $\mu^*$ on $\alphanut g$ for the Poisson, uniform ($r=0.9,0.99,0.999,1$), and power-law distributions for $\nu=2$ and 3, respectively. The growth rate $\mu^*$ is smallest for the Poisson out-degree distribution and largest for the power-law out-degree distribution, suggesting that network heterogeneity enhances cell growth under nutrient-poor conditions. Indeed, for the uniform distribution, $\mu^*$ increases as the heterogeneity parameter $r$ increases. As in the case of $\nu=1$, network heterogeneity also helps prevent starvation. As shown in Figs.~\ref{fig:starvation}(c) and (d), for the Poisson out-degree distribution, $\mu^*$ becomes negative when $\alphanut g<1-\alphacat$ (see also Fig.~\ref{fig:schematic_dyna}), whereas for the uniform distribution, the positive-$\mu^*$ region expands as $r$ increases, with $\mu^*$ remaining positive for any $\alphanut g$ at $r=1$. For the power-law distribution, $\mu^*$ also remains positive for any $\alphanut g$, and the metabolic--starvation transition is absent.

Assuming $\mu^* \ll \alphanut g \ll 1$, we can obtain the asymptotic forms of $\mcat^*, m_{\rm in}^*$, and $\mu^*$ for the uniform distribution with $r=1$ and the power-law distribution (see Appendix Sec.~\ref{sec:asymptotic} for a detailed derivation). For both distributions, $\mcat^*$ converges to $(1-\alphanut g)/\alphacat$. Using this relation in Eq.~\eqref{eq:mcat1} and evaluating $X(\mu^*, (\mcat^*)^\nu)$, we obtain $\mu^* = 2(\mcat^*)^\nu \exp(-\frac{2(1-\alphanut g)}{\alphacat m_{\rm in}^*})$ for the uniform distribution with $r=1$ and $\mu^* = (\mcat^*)^\nu \exp(-\frac{3}{2}-\frac{1-\alphanut g}{2\alphacat m_{\rm in}^*})$ for the power-law distribution. Moreover, when $\mu^* \ll \alphanut g \ll 1$, $m_{\rm in}^*$ is proportional to $\alphanut g$; hence, $\mu^*$ takes an essential singularity form of $\exp(-C/\alphanut g)$ irrespective of $\nu$, where $C$ is a positive constant. These asymptotic solutions are also shown in Figs.~\ref{fig:starvation} (c) and (d). 

Mathematically, these asymptotic behaviors originate from the logarithmic divergence of $X(\mu^*, (\mcat^*)^\nu)$ as $\mu^* \to 0$. This divergence, in turn, originates from the divergence of $x_{{\rm cat},uv}^*$, given by Eq.~\eqref{eq:fixed_cat}, as $v/c \to 0$. Thus, the presence or absence of Phase S is determined by whether the network contains catalytic chemical species with extremely small out-degrees, for which the outgoing reaction flux is negligible compared with the incoming flux. This mechanism is therefore independent of $\nu$.


\begin{figure}[t]
\centering
\includegraphics[width=85mm]{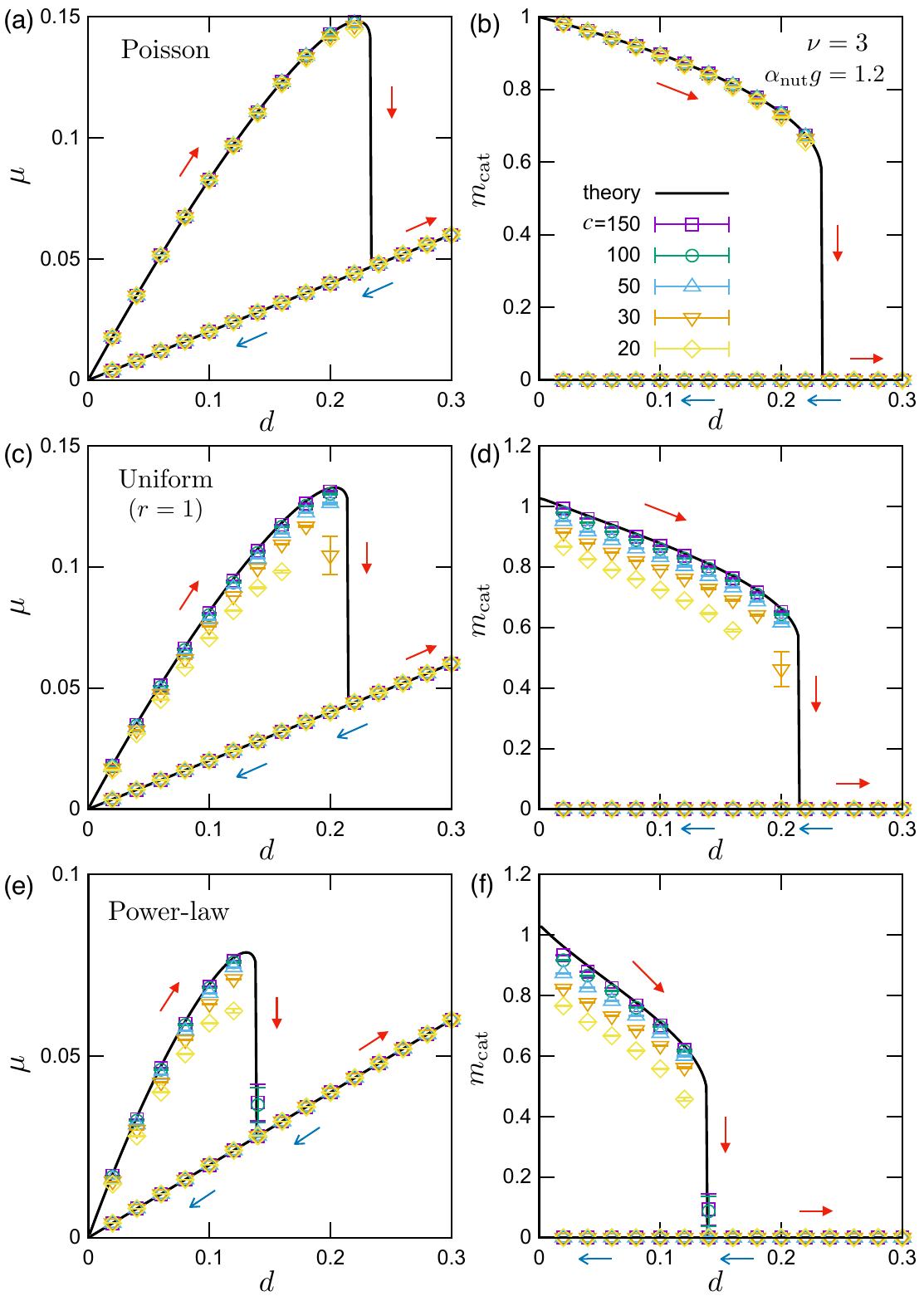}
\caption{The growth rate $\mu$ and the average abundance of catalytic chemical species $\mcat$ against the permeability $d$ for $\nu = 3$ and $\alphanut g = 1.2$; (a) $\mu$ and (b) $\mcat$ with a Poisson out-degree distribution; (c) $\mu$ and (d) $\mcat$ with a uniform out-degree distribution ($r=1$); (e) $\mu$ and (f) $\mcat$ with a power-law out-degree distribution. The black curve represents $\mu^*$ and $\mcat^*$ obtained from the DMFT. The data points are obtained from numerical simulations with finite connectivities $c = 150, 100, 50, 30, 20$ and $N=10^5$. The simulation data are averaged over 10 network realizations. In the simulations, $d$ is swept upward and then downward, as indicated by the red and blue arrows, respectively. At each value of $d$, the system is evolved for $t=400$, and $\mu(t)$ and $\mcat (t)$ at the final state are plotted, and the final state is then used as the initial state for the next value of $d$. 
}
\label{fig:hys_rich}
\end{figure}

\begin{figure}[t]
\centering
\includegraphics[width=85mm]{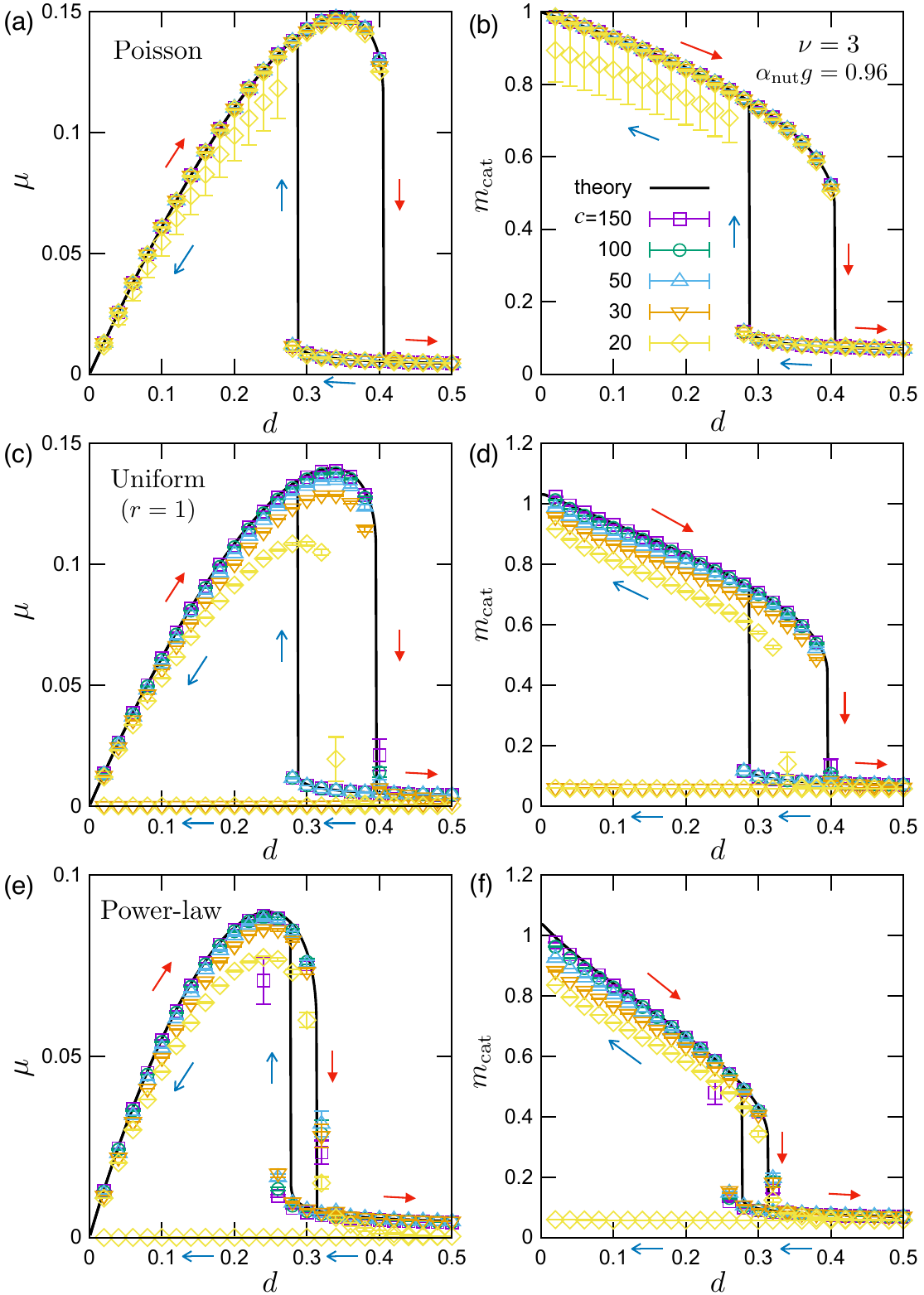}
\caption{
The growth rate $\mu$ and the average abundance of catalytic chemical species $\mcat$ against the permeability $d$ for $\nu = 3$ and $\alphanut g = 0.96$; (a) $\mu$ and (b) $\mcat$ with a Poisson out-degree distribution; (c) $\mu$ and (d) $\mcat$ with a uniform out-degree distribution ($r=1$); (e) $\mu$ and (f) $\mcat$ with a power-law out-degree distribution (exponent $\beta = 3$). The black curve represents $\mu^*$ and $\mcat^*$ obtained from the DMFT. The data points are obtained from numerical simulations with finite connectivities $c = 150, 100, 50, 30, 20$ and $N=10^5$. The simulation data are averaged over 10 network realizations. In the simulations, $d$ is swept upward and then downward, as indicated by the red and blue arrows, respectively. At each value of $d$, the system is evolved for $t=400$, and $\mu(t)$ and $\mcat (t)$ at the final state are plotted, and the final state is then used as the initial state for the next value of $d$. 
}
\label{fig:hysteresis}
\end{figure}

\section{Bistability}
\label{sec:bistability}

We next examine the behavior of physical quantities under high nutrient supply. For $\nu=2$ and $3$, Phase~II, in which the overnutrition and metabolic states coexist, appears in the region $\alphanut g>1$ (see Fig.~\ref{fig:phase}). For $\nu=3$, Phase~III also appears in the region $\alphanut g<1$, where low- and high-growth metabolic states coexist. In the presence of such bistable phases, hysteretic behavior can arise when the parameters are varied dynamically. To examine this hysteresis, we perform numerical simulations of Eq.~\eqref{eq:dynamical1} on networks with Poisson, uniform ($r=1$), and power-law out-degree distributions at finite mean connectivities $c=150$, $100$, $50$, $30$, and $20$, with $N=1.0\times10^5$. Starting from a small value of $d$, we gradually increase $d$, relaxing the system at each $d$, and then decrease $d$ in the same manner (see Appendix Sec.~\ref{sec:simulation_methods} for details of the numerical procedure).

Figures~\ref{fig:hys_rich}(a) and (b) show the hysteretic behaviors of the cell growth rate $\mu$ and $\mcat$ at $\alphanut g=1.2$ for the Poisson out-degree distribution for $\nu=3$. For $c\geq20$, the simulation results are in good agreement with the DMFT predictions. During the $d$-increasing process, the system remains in the metabolic state until it suddenly jumps to the overnutrition state. In contrast, during the $d$-decreasing process, the system remains in the overnutrition state and does not return to the metabolic state even when $d$ is decreased to zero.  This is because, for $\nu\geq2$, the overnutrition state remains stable for any $d$ whenever $\alphanut g>1$ (see Appendix Sec.~\ref{sec:stability}). As shown in Figs.~\ref{fig:hys_rich}(c)--(f), the uniform and power-law out-degree distributions exhibit qualitatively the same hysteretic behavior, although Phase~II is narrower than in the Poisson case. However, $\mu$ and $\mcat$ gradually shift downward from the DMFT predictions as $c$ decreases. Finite-connectivity effects also shift the phase boundary: for the uniform distribution at $c=20$, the metabolic state is destabilized at a smaller value of $d$ than predicted by DMFT, as shown in Figs.~\ref{fig:hys_rich}(c) and (d).

Figures~\ref{fig:hysteresis}(a) and (b) show the hysteretic behavior of the cell growth rate $\mu$ and $\mcat$ at $\alphanut g=0.96$ for $\nu=3$ with the Poisson out-degree distribution. During the $d$-increasing process, the system transitions from the high-growth to the low-growth metabolic state. During the $d$-decreasing process, the reverse transition occurs at a smaller value of $d$, resulting in a hysteresis loop. Finite-connectivity effects are weak for the Poisson distribution for $c\ge 30$. However, for $c=20$, the transition from the low-growth to the high-growth metabolic state does not occur even down to $d=0$ for some network realizations, suggesting that network sparsity suppresses the destabilization of the low-growth metabolic state.

This finite-connectivity effect becomes more pronounced for the uniform and power-law out-degree distributions. As shown in Figs.~\ref{fig:hysteresis}(c)--(f), both distributions exhibit smaller hysteresis loops than the Poisson distribution, consistent with the narrower Phase~III regions shown in Figs.~\ref{fig:phase} (f) and (i). As observed for $\alphanut g=1.2$, the simulation results deviate increasingly from the theoretical curves as $c$ decreases during the $d$-increasing process. Moreover, for the uniform distribution at $c=30$ and $20$ and for the power-law distribution at $c=20$, the low-growth metabolic state persists even when $d$ is decreased to zero, indicating a stronger stabilization of the low-growth metabolic state due to network sparsity.


\section{Summary and Discussion}
\label{sec:summary}

In this study, we generalized the Furusawa--Kaneko model to multicatalyst reactions, in which each reaction is catalyzed by $\nu$ chemical species, and analyzed the resulting dynamics using DMFT. In the dense-connectivity and thermodynamic limits, we derived effective dynamical equations applicable to arbitrary degree distributions and examined Poisson, uniform, and power-law out-degree distributions. From the fixed-point analysis, we find the phase behavior only depends on the out-degree distribution. Under nutrient-poor conditions, network heterogeneity suppresses starvation irrespective of $\nu$, and the starvation phase disappears for the power-law out-degree distribution. Under nutrient-rich conditions, multicatalyst reactions qualitatively change the bifurcation structure: for $\nu\geq2$, a bistable phase emerges in which the metabolic and overnutrition states coexist, accompanied by a discontinuous transition, while for $\nu\geq3$, an additional bistable phase appears in which two metabolic states with different cell growth rates coexist, leading to discontinuous switching between the two states. These qualitative phase structures are common to the Poisson, uniform, and power-law out-degree distributions. Numerical simulations at finite mean connectivity further showed that the DMFT predictions provide a good description of the finite-connectivity systems, while deviations become more pronounced as the network becomes sparse.

For the Poisson out-degree distribution, the origin of the discontinuous transitions can be understood directly from the reduced dynamics of $\mcat$. Because the dynamical equation closes in terms of $\mcat$ alone, the bifurcation structure can be classified comprehensively from the shape of $f_\nu(\mcat)=\mcat^{\nu-1}(1-\mcat)$, in particular, from the numbers of its extrema and inflection points. A notable feature is that the original model with triadic interactions ($\nu=1$) exhibits only continuous transitions, whereas tetradic interactions ($\nu=2$) give rise to bistability and discontinuous transitions. For pentadic interactions ($\nu=3$), the bifurcation structure changes further, leading to the coexistence of two distinct metabolic states with different growth rates. For the uniform and power-law out-degree distributions, the dynamics cannot be reduced to a single equation for $\mcat$. Nevertheless, their qualitative phase behavior remains the same as in the Poisson case. This robustness may be attributed to the nonlinear factor $(\mcat^*)^\nu$ appearing in Eqs.~\eqref{eq:mcat1} and \eqref{eq:mnut2}, which governs the number of fixed-point solutions. The qualitative changes in phase behavior induced by increasing the interaction order may have broader implications beyond multicatalyst reaction networks. Recent studies of gene-regulatory and ecological networks have often modeled higher-order effects through triadic interactions~\cite{niedostatek2025mining, bairey2016high-order, grilli2017higher-order}; our results suggest that extending such descriptions to tetradic and pentadic interactions may capture even richer dynamical phenomena. 

Finally, several issues should be noted although our results may provide useful insights into real metabolic networks. First, the Furusawa--Kaneko model is a coarse-grained model, and its dynamical variables do not correspond one-to-one to individual chemical species in real cells, such as mRNAs and metabolites. Our multicatalyst reactions are motivated by linear reaction pathways in which successive reactions are catalyzed by different enzymes, with the enzymes potentially organized in multienzyme complexes~\cite{srere1987complexes, kohnhorst2017identification, pareek2021metabolic}. We phenomenologically represent such pathways by a single multicatalyst reaction. How this effective description can be derived microscopically and whether it can generate bistability and abrupt switching in real metabolic networks remain an open problem. Another important limitation concerns network sparsity. Real metabolic networks are sparse and are reported to have a mean connectivity of typically $c=3$~\cite{jeong2000the}. In contrast, our exact analysis is performed in the dense-connectivity limit, and our numerical simulations were performed for $c\geq20$, which is still considerably denser than real metabolic networks. Although the discontinuity appears to persist at finite connectivity, obvious finite-connectivity effects are observed even for $c\geq20$. The effect of network sparsity on the phase behavior therefore requires further systematic study. Despite these limitations, exact analysis of a simple model provides a useful theoretical baseline for understanding more realistic metabolic networks.

 \begin{acknowledgments}
This work was supported by Japan Society for the Promotion of Science KAKENHI (Grant Nos. 25K17354 and 24K00594, K.M.; Grant No. 24H01931 and 26H00951, S.I.)
 \end{acknowledgments}

\appendix
\onecolumngrid

\section{Simulation methods} 
\label{sec:simulation_methods}
We specify the multicatalyst reaction network by a binary adjacency tensor $\cijk \in \{0,1\}$. The tensor $\cijk$ takes the value 1 if the reaction $\reaction$ exists, and 0 otherwise. The network used in the numerical simulations is generated in three steps. First, we construct a directed network described by an adjacency matrix $A_{ij} \in \{0,1\}$, where $A_{ij}=1$ indicates that the reaction channel $j\to i$ is allowed. Second, we divide the nodes into two groups according to the type of chemical species they represent: catalytic species, $\mathcal{G}{\rm cat}$, and nutrients, $\mathcal{G}{\rm nut}$. The fractions assigned to these groups are fixed to be $\alphacat$ and $\alphanut$, respectively, with $\alphacat+\alphanut = 1$. Finally, for each directed edge with $A_{ij}=1$, we independently assign a tuple of catalysts $(k_1,k_2,\ldots,k_\nu)$ with each $k_\ell$ chosen uniformly at random from $\mathcal{G}_{\rm cat}$~\cite{furusawa2003zipf, mitsumoto2026starvation}. 

For the Poisson out-degree distribution, we employ an Erd\H{o}s-R\'{e}nyi random network. $A_{ij}$ is set to 1 with probability $c/N$ and to $0$ otherwise~\cite{erdos1959random, barabasi2016network, newman2010networks}. For the uniform and power-law out-degree distributions, we use a configuration model~\cite{chung2002connected, lee2009critical, kim2026heterogeneous}. For the uniform case, the in- and out-degree distributions, $\pin$ and $\pout$, are both taken to be uniform over the interval $[(1-r)c,(1+r)c]$. For the power-law case, we take $\pin = \delta_{u,c}$ and $\pout = 2c(c+1)/(v+c)(v+c+1)(v+c+2)$. We independently draw an in-degree $u_i$ and an out-degree $v_i$ for each node $i$ from $\pin$ and $\pout$, respectively, and then determine each element of the adjacency matrix according to 
$P(A_{ij})=\frac{u_i v_j}{Nc}\delta_{A_{ij},1}+\qty(1-\frac{u_i v_j}{Nc})\delta_{A_{ij},0}$.

For the numerical simulation of Eq.~\eqref{eq:dynamical1} for $\nu=3$, we set the initial abundance of all chemical species to $x_i=1$. Starting from $d=0.02$, we evolve the system for $400$ time units to relax. We then increase $d$ by $0.02$, use the resulting state as the initial condition, and evolve the system for another $400$ time units. This procedure is repeated until $d$ reaches $0.5$. Subsequently, starting from the final state at $d=0.5$, we decrease $d$ by $0.02$ at each step and repeat the same relaxation procedure until $d$ reaches $0.02$. We solve Eq.~\eqref{eq:dynamical1} by using the fourth-order Runge-Kutta method. The time resolution $\Delta t = 0.02$ is used. 

\section{Nondimensionalization of dynamical equation}
\label{sec:nondimension}
We present the detailed description of the dynamical equation, Eq.~\eqref{eq:dynamical1}, and the nondimensionalization of the variables and parameters.
Let $n_i$ be the number of molecules of chemical species $i$ in a cell of volume $V~[\mathrm{L}^3]$, and let $\rho_i=n_i/V~[\mathrm{L}^{-3}]$ be its concentration. The time derivative of $\rho_i$ is written as
\begin{align}
\frac{d\rho_i}{dt} = \frac{1}{V}\frac{dn_i}{dt}-\rho_i \frac{1}{V}\frac{dV}{dt}.
\end{align}
The first term on the right-hand side represents changes in the molecule number due to multicatalyst reactions and exchange with the external environment. We write it as
\begin{align}
\frac{1}{V}\frac{dn_i}{dt} &= \eta \sum_{j,\{k\}} \cijk \rho_j\rho_{k_1}\cdots \rho_{k_\nu} -\eta \sum_{j,\{k\}} \cjik\rho_i\rho_{k_1}\cdots \rho_{k_\nu}+d_i(g_i-\rho_i),
\label{eq:chem_reaction}
\end{align}
where $\eta~[\mathrm{L}^{3\nu} \mathrm{T}^{-1}]$ is the reaction rate constant,
$\cijk \in\{0,1\}$ is a binary tensor specifying whether the reaction $\reaction$ exists, $d_i~[\mathrm{T}^{-1}]$ is the exchange rate with the external environment, and $g_i~[\mathrm{L}^{-3}]$ is the equilibrium concentration outside the cell. We define the cell growth rate as $\mu=(1/V)dV/dt~[\mathrm{T}^{-1}]$. The concentration dynamics before nondimensionalization is therefore
\begin{align}
\frac{d\rho_i}{dt}&=\eta \sum_{j,\{k\}} \cijk\rho_j\rho_{k_1}\cdots \rho_{k_\nu} -\eta \sum_{j,\{k\}} \cjik\rho_i\rho_{k_1}\cdots \rho_{k_\nu} +d_i(g_i-\rho_i)-\mu \rho_i.
\label{eq:dimensional_equation}
\end{align}
Let $N_{\rm tot}$ be the total number of molecules in the cell, i.e., $N_{\rm tot}=\sum_{i=1}^N n_i$. By taking the sum of Eq. \eqref{eq:chem_reaction}, one finds 
\begin{align}
\frac{dN_{\rm tot}}{dt}=V\sum_{i=1}^N d_i(g_i-\rho_i).
\end{align}
Now, we employ a minimal description of the cell-volume regulation that the total molecular concentration, $N_{\rm tot}/V$, remains constant in time. This assumption implies
\begin{align}
    \frac{1}{V}\frac{dV}{dt}=\frac{1}{N_{\rm tot}}\frac{dN_{\rm tot}}{dt}
\label{eq:relation_V_Ntot}
\end{align}
as follows from $\frac{d}{dt}\log(N_{\rm tot}/V)=0$. Using Eq.~\eqref{eq:relation_V_Ntot}, the growth rate is obtained as
\begin{align}
    \mu=\frac{V}{N_{\rm tot}}\sum_{i=1}^N d_i(g_i-\rho_i).
\end{align}
Thus, maintaining a constant total molecular concentration determines the growth rate as given above.

Let us introduce a dimensionless abundance $x_i = \rho_i/\rho_0$, where $\rho_0=N_{\rm tot}/VN$ is the concentration scale, to satisfy the normalization given by Eq.~\eqref{eq:constraint1}. We also nondimensionalize time as $\tilde{t} = c\eta\rho_0^\nu t$, where $c$ is the mean connectivity of the reaction network. Substituting these rescaled variables into the dimensional concentration dynamics, we obtain the dimensionless dynamical equation as
\begin{align}
\frac{dx_i}{d\tilde{t}} &= \frac{1}{c}\sum_{j,\{k\}}C_{ij}^{k_1\ldots k_\nu} x_j x_{k_1}\cdots x_{k_\nu} - \frac{1}{c}\sum_{j,\{k\}}C_{ji}^{k_1\ldots k_\nu} x_i x_{k_1}\cdots x_{k_\nu} + \tilde{d}_i(\tilde{g}_i-x_i) - \tilde{\mu}x_i,
\end{align}
where $\tilde{d}_i = d_i/c\eta\rho_0^\nu$, $\tilde {g}_i = g_i/\rho_0$, and $\tilde{\mu} = \mu/c\eta\rho_0^\nu = \frac{1}{N}\sum_{i=1}^N \tilde{d}_i(\tilde {g}_i-x_i)$ are dimensionless permeability, equilibrium abundance outside the cell, cellular growth rate, respectively. Because we have assumed that $N_{\rm tot}/V$ is constant, $\tilde{d}_i$ and $\tilde{g}_i$ can be treated as constant parameters. In the main text, we denote them simply as $d_i$, $g_i$, and $\mu$, respectively.

\section{Derivation of the dynamical mean-field equation}
\label{sec:dmft_derivation}

We consider a multicatalyst reaction network model consisting of $N$ chemical species~\cite{furusawa2003zipf, furusawa2006evolutionary, mitsumoto2026starvation}. The chemical species are divided into catalytic chemical species and nutrients, denoted by $\mathcal{G}_{\rm cat}$ and $\mathcal{G}_{\rm nut}$, respectively. The fractions of chemical species belonging to these groups are denoted by $\alphacat$ and $\alphanut$, respectively, satisfying $\alphacat + \alphanut = 1$. The model dynamics are described by the following differential equations:
\begin{align}
\dot{x}_i(t) &= \frac{1}{c}\sum_{j =1}^N \sum_{\{k\} \in \mathcal{G}_{\rm cat}} \cijk x_j(t) x_{k_1}(t)\cdots x_{k_\nu}(t) \nonumber \\
&-  \frac{1}{c}\sum_{j =1}^N \sum_{\{k\} \in \mathcal{G}_{\rm cat}}\cjik x_i(t) x_{k_1}(t)\cdots x_{k_\nu}(t) + d_i(g_i-x_i(t)) - \mu(t) x_i(t) + h_i(t) + \xi_i(t),
\label{eq:SM_model}
\end{align}
where the parameters $(d_i, g_i)$ are defined as $(0, 0)$ for $i \in \mathcal{G}_{\rm cat}$ and $(d, g)$ for $i \in \mathcal{G}_{\rm nut}$. $\mu = \frac{1}{N}\sum_{i =1}^N d_i(g_i-x_i)$ denotes the cell growth rate, and the term $- \mu(t) x_i(t)$ represents the dilution of chemical abundances due to the cell growth. The site-dependent field $h_i(t)$ is introduced for theoretical convenience and is set to zero at the end of the calculation. The variable $\xi_i(t)$ represents a white Gaussian noise with zero mean and variance $\eta^2$ (in the main text, we present results for $\xi_i(t) = 0$). The tensor $\cijk$ is equal to 1 if the reaction $\reaction$ exists, and is equal to 0 otherwise. We suppose that the nutrients cannot act as catalysts, i.e., $\cijk = 0$ if any $k$ belongs to $\mathcal{G}_{\rm nut}$. $\cijk$ is generated using the configuration model~\cite{chung2002connected, newman2010networks}. Specifically, we first draw the in-degree $u_i$ and out-degree $v_i$ for all nodes $i$ independently from the distributions $\pin$ and $\pout$. The probability distribution of $\cijk$ is then given by
\begin{align}
P(\cijk) = \frac{u_iv_j}{(N\alphacat)^\nu Nc}\delta_{\cijk,1} + \qty(1-\frac{u_iv_j}{(N\alphacat)^\nu Nc})\delta_{\cijk,0},
\label{eq:SM_cijk}
\end{align}
where $c$ is the mean connectivity, i.e., $\lim_{N\to\infty}\frac{1}{N}\sum_{i,j,\{k\}}\cijk = \sum_{u}\pin u = \sum_{v}\pout v = c$. The factor $1/(N\alphacat)^{\nu}$ accounts for the random selection of the $\nu$ catalysts from the $N\alphacat$ catalytic chemical species in $\mathcal{G}_{\rm cat}$. 

We solve the dynamics of Eq.~\eqref{eq:SM_model} on a network generated by  Eq.~\eqref{eq:SM_cijk} using dynamical mean-field theory~\cite{martin1973statistical, janssen1976on, dedominicis1978dynamics, sompolinsky1982relaxational, cugliandolo1993analytical, castellani2005spin, galla2024generating}. The theory is based on the generating functional
\begin{align}
\mathcal{Z}[\bm{h},\bm{\psi}] = \int \qty(\prod_{i=1}^N Dx_i) \mP{\bm{x}, \bm{h}}e^{i\int dt \sum_{i=1}^N x_i(t)\psi_i(t)}.
\label{eq:SM_GF_original}
\end{align}
The path probability distribution of $\bm{x}$ can be written as
\begin{align}
\mP{\bm{x}, \bm{h}} = &P_0(\bm{x}(0)) \int \qty(\prod_{i=1}^N D\xi_i)\mathcal{P}_{\eta}[\bm{\xi}] \nonumber \\
&\times\delta_{\rm F} \Bigg[\dot{x}_i(t)-\frac{1}{c}\sum_{j,\{k\}}\cijk x_j(t)x_{k_1}(t)\cdots x_{k_\nu}(t) - \frac{1}{c}\sum_{j,\{k\}} \cjik x_i(t)x_{k_1}(t)\cdots x_{k_\nu}(t) \nonumber \\
&-d_i(g_i-x_i(t))+\mu(t)x_i(t)-h_i(t)-\xi_i(t)\Bigg],
\label{eq:SM_path_original}
\end{align}
where $\mathcal{P}_{\eta}[\bm{\xi}]$ is the path probability distribution of the Gaussian noise
\begin{align}
\mathcal{P}_{\eta}[\bm{\xi}] \sim \exp\qty(-\frac{1}{2\eta^2}\int dt \sum_{i=1}^N\xi_i^2(t))
\label{eq:SM_path_noise}
\end{align}
and $P_0(\bm{x}(0))$ represents the joint initial distribution for all nodes $i$.
Using a functional integral representation of $\delta_{\rm F}$ and integrating out the Gaussian noise $\bm{\xi}_i$, we obtain
\begin{align}
\mathcal{Z}[\bm{h},\bm{\psi}] = \int \qty(\prod_{i=1}^N Dx_i D\hat{x}_i) P_0(\bm{x}(0))&\exp\Bigg(-\frac{i}{c}\sum_{i,j,\{k\}}\cijk \int dt \hat{x}_i(t) x_j(t) x_{k_1}(t)\cdots x_{k_\nu}(t) \nonumber \\
&~~~~~~~~+ \frac{i}{c}\sum_{i,j,\{k\}}\cjik \int dt \hat{x}_i(t) x_i(t)x_{k_1}(t)\cdots x_{k_\nu}(t)\Bigg) \nonumber \\
\times &\exp\qty(i\int dt \sum_{i=1}^N A_i(x_i(t), \hat{x}_i(t), h_i(t), \psi_i(t))),
\label{eq:SM_GF1}
\end{align}
where a single-site action is defined as
\begin{align}
A_i(x_i(t), \hat{x}_i(t), h_i(t), \psi_i(t)) = x_i(t)\psi_i(t)+\frac{i \eta^2}{2}\hat{x}_i^2(t) + \hat{x}_i(t)[\dot{x}_i(t)-d_i(g_i-x_i(t))+\mu(t)x_i(t)-h_i(t)].
\label{eq:SM_action}
\end{align}
From Eqs.~\eqref{eq:SM_GF1} and \eqref{eq:SM_action}, one finds that the fields $h_i(t)$ and $\psi_i(t)$ are coupled to the dynamical variables $\hat{x}_i(t)$ and $x_i(t)$, respectively. This leads to the following formulae,
\begin{align}
\expval{x_i(t)}_\xi &= (-i) \lim_{\bm{h}, \bm{\psi}\to 0} \frac{\delta Z[\bm{h}, \bm{\psi}]}{\delta \psi_i(t)}, \label{eq:SM_moment}  \\
\expval{\hat{x}_i(t)}_\xi &= i \lim_{\bm{h}, \bm{\psi}\to 0} \frac{\delta Z[\bm{h}, \bm{\psi}]}{\delta h_i(t)} = 0, \label{eq:SM_moment_hat}  \\
R(t_1,t_2) &= \frac{1}{N}\sum_{i=1}^N \lim_{\psi_i\to 0} \frac{\delta \expval{x_i(t_2)}_\xi}{\delta h_i(t_1)} = (-i) \lim_{\bm{h}, \bm{\psi}\to 0} \frac{1}{N}\sum_{i=1}^N\frac{\delta^n Z[\bm{h}, \bm{\psi}]}{\delta h_i(t_1) \delta \psi_i(t_2)} = \frac{-i}{N}\sum_{i=1}^N \expval{\hat{x}_i(t_1)x_i(t_2)}_\xi, \label{eq:SM_response} 
\end{align}
where $\expval{\cdots}_\xi$ denotes the average over the noise and $R(t_1,t_2)$ represents the response function. Because of the causality, $R(t_1,t_2) = 0$ if $t_2 \le t_1$, although $\lim_{t_2 \to t_1^-}R(t_1,t_2) \neq 0$.
We also note that the generating functional satisfies the normalization condition $Z[\bm{h}, 0] = 1$, leading to $\expval{\hat{x}_i(t)}_\xi=0$, as given in Eq.~\eqref{eq:SM_moment_hat}.

Let us evaluate the generating functional $\ovl{\mathcal{Z}[\bm{h},\bm{\psi}]}$ averaged over the network topology $\qty{\cijk}$. In general, this averaged generating functional can be written as
\begin{align}
\ovl{\mathcal{Z}[\bm{h},\bm{\psi}]} = \int \qty(\prod_{i=1}^N Dx_i D\hat{x}_i) P_0(\bm{x}(0)) \Delta [\bm{x}, \hat{\bm{x}}] \exp\qty(i\int dt \sum_{i=1}^N A_i(x_i(t), \hat{x}_i(t), h_i(t), \psi_i(t))),
\label{eq:SM_GF2}
\end{align}
where
\begin{align}
\Delta [\bm{x}, \hat{\bm{x}}] &= \ovl{ \exp\qty(-\frac{i}{c}\sum_{i,j,\{k\}} \cijk \int dt \hat{x}_i(t) x_j(t)x_{k_1}(t)\cdots x_{k_\nu}(t) + \frac{i}{c}\sum_{i,j,\{k\}}\cijk \int dt \hat{x}_i(t) x_i(t)x_{k_1}(t)\cdots x_{k_\nu}(t))} \notag \\
&= \ovl{ \prod_{i,j,\{k\}} \exp\qty(-\frac{i}{c}\cijk \int dt  ( \hat{x}_i(t) -  \hat{x}_j(t))x_j(t)x_{k_1}(t)\cdots x_{k_\nu}(t))} \nonumber \\
&= \exp \qty[\sum_{i,j,\{k\}} \ln \qty(1+\frac{u_iv_j}{(N\alphacat)^\nu Nc}\qty(\exp \qty( -\frac{i}{c}\int dt ( \hat{x}_i(t) -  \hat{x}_j(t)) x_j(t)x_{k_1}(t)\cdots x_{k_\nu}(t))-1))].
\label{eq:SM_delta1}
\end{align}
To decouple the interaction terms between different sites in $\Delta [\bm{x}, \hat{\bm{x}}]$, we introduce the following functional order parameters~\cite{mimura2009parallel, poley2025interaction, metz2025dynamical}: 
\begin{align}
U_{{\rm cat},u}[\hat{x}] &= \frac{1}{\alphacat \pin N}\sum_{i \in \mathcal{G}_{\rm cat}} \delta_{u_i,u} \delta_{\rm F}[\hat{x}(t)-\hat{x}_i(t)], \label{eq:SM_order1_cat} \\
U_{{\rm nut},u}[\hat{x}] &= \frac{1}{\alphanut \pin N}\sum_{i \in \mathcal{G}_{\rm nut}} \delta_{u_i,u} \delta_{\rm F}[\hat{x}(t)-\hat{x}_i(t)], \label{eq:SM_order1_nut} \\
V_{{\rm cat},v}[x, \hat{x}] &= \frac{1}{\alphacat \pout N}\sum_{i \in \mathcal{G}_{\rm cat}} \delta_{v_i,v} \delta_{\rm F}[x(t)-x_i(t)]\delta_{\rm F}[\hat{x}(t)-\hat{x}_i(t)], \label{eq:SM_order2_cat}\\
V_{{\rm nut},v}[x, \hat{x}] &= \frac{1}{\alphanut \pout N}\sum_{i \in \mathcal{G}_{\rm nut}} \delta_{v_i,v} \delta_{\rm F}[x(t)-x_i(t)]\delta_{\rm F}[\hat{x}(t)-\hat{x}_i(t)], \label{eq:SM_order2_nut}\\
W[x] &= \frac{1}{\alphacat N}\sum_{i \in \mathcal{G}_{\rm cat}}\delta_{\rm F}[x(t)-x_i(t)].\label{eq:SM_order3}
\end{align}
For notational convenience, we introduce $U_{\sigma,u}$ and $V_{\sigma,v}$ with $\sigma = {\rm cat},{\rm nut}$.
Using these order parameters, we can write $\Delta [\bm{x}, \hat{\bm{x}}]$ as
\begin{align}
\Delta [\bm{x}, \hat{\bm{x}}] = \exp \Bigg[N \sum_{u,v}\pin \pout &\int D\hat{x} Dy D\hat{y} Dz_1\cdots Dz_\nu(\sum_{\sigma} \alpha_\sigma U_{\sigma,u}[\hat{x}])(\sum_{\sigma} \alpha_\sigma V_{\sigma,v}[y, \hat{y}]) W[z_1]\cdots W[z_\nu] f_{uv}[\hat{x}, y, \hat{y}, \{z\}]\Bigg],
\label{eq:SM_delta2}
\end{align}
where
\begin{align}
f_{uv}[\hat{x}, y, \hat{y}, \{z\}] = N(N\alphacat)^\nu \ln \qty(1+\frac{uv}{(N\alphacat)^\nu Nc}\qty(\exp \qty(-\frac{i}{c}\int dt (\hat{x}(t) -  \hat{y}(t))y(t) z_1(t)\cdots z_\nu(t))-1)).
\label{eq:SM_fuvw1}
\end{align}
In the dense connectivity limit, the leading terms in powers of $1/c$ are relevant. Hence, by expanding the exponential in Eq.~\eqref{eq:SM_fuvw1} with respect to $1/c$ and taking the thermodynamic limit $N\to\infty$ and, we obtain
\begin{align}
f_{uv}[\hat{x}, y, \hat{y}, \{z\}] = -i\frac{uv}{c^2}\int dt (\hat{x}(t) -  \hat{y}(t))y(t) z_1(t)\cdots z_\nu(t).
\label{eq:SM_fuvw2}
\end{align}
By inserting the following identities,
\begin{align}
1 &= \int DU_{\sigma, u} \delta_{\rm F}\Bigl[\alpha_\sigma \pin N U_{\sigma, u}[\hat{x}]-\sum_{i \in \mathcal{G}_\sigma}\delta_{u_i,u} \delta_{\rm F}[\hat{x}(t)-\hat{x}_i(t)]\Bigr] \nonumber \\
&= \frac{1}{2\pi} \int DU_{\sigma, u} D\hat{U}_{\sigma, u} e^{i\alpha_\sigma \pin N\int D\hat{x} \hat{U}_{\sigma, u}[\hat{x}] U_{\sigma, u}[\hat{x}]-i\sum_{i \in \mathcal{G}_\sigma} \delta_{u_i,u}\hat{U}_{\sigma, u}[\hat{x}_i]}, \label{eq:SM_identity1} \\
1 &= \int DV_{\sigma, v} \delta_{\rm F}\Bigl[\alpha_\sigma \pout NV_{\sigma, v}[x,\hat{x}]-\sum_{i \in \mathcal{G}_\sigma}\delta_{v_i,v} \delta_{\rm F}[x(t)-x_i(t)]\delta_{\rm F}[\hat{x}(t)-\hat{x}_i(t)]\Bigr] \nonumber \\
&= \frac{1}{2\pi} \int DV_{\sigma, v} D\hat{V}_{\sigma, v} e^{i \alpha_\sigma \pout N\int Dx D\hat{x} \hat{V}_{\sigma, v}[x, \hat{x}] V_{\sigma, v}[x, \hat{x}]-i\sum_{i \in \mathcal{G}_\sigma}\delta_{v_i,v}\hat{V}_{\sigma, v}[x_i, \hat{x}_i]}, \label{eq:SM_identity2} \\
1 &= \int DW_{\rm cat} \delta_{\rm F}\Bigl[\alphacat NW_{\rm cat}[x]-\sum_{i \in \mathcal{G}_{\rm cat}} \delta_{\rm F}[x(t)-x_i(t)]\Bigr] \nonumber \\
&= \frac{1}{2\pi}\int DW D\hat{W} e^{i \alphacat N\int Dx \hat{W}[x] W[x]-i\sum_{i \in \mathcal{G}_{\rm cat}} \hat{W}[x_i]},
\end{align}
into the averaged generating functional $\ovl{Z[\bm{h},\bm{\psi}]}$, We obtain
\begin{align}
\ovl{\mathcal{Z}[\bm{h},\bm{\psi}]} = \int \qty(\prod_{\sigma,u}DU_{\sigma, u}D\hat{U}_{\sigma, u})\qty(\prod_{\sigma,v}DV_{\sigma, v}D\hat{V}_{\sigma, v})DWD\hat{W}\exp(N\Phi[\bm{U},\hat{\bm{U}},\bm{V}, \hat{\bm{V}}, W, \hat{W}]),
\label{eq:SM_GF3}
\end{align}
where
\begin{align}
\Phi &= \sum_{u,v}\pin \pout \int D\hat{x} Dy D\hat{y} Dz_1\cdots Dz_\nu (\sum_{\sigma} \alpha_\sigma U_{\sigma, u}[\hat{x}])(\sum_{\sigma} \alpha_\sigma V_{\sigma, v}[y, \hat{y}]) W[z_1]\cdots W[z_\nu]f_{uv}[\hat{x}, y, \hat{y}, \{z\}] \nonumber \\
&+ i\sum_{\sigma} \alpha_\sigma \qty(\sum_u \pin \int D\hat{x} U_{\sigma, u}[\hat{x}]\hat{U}_{\sigma, u}[\hat{x}] + \sum_v \pout \int Dx D\hat{x} V_{\sigma, v}[x, \hat{x}]\hat{V}_{\sigma, v}[x, \hat{x}]) + \nu\alphacat\int Dx W[x]\hat{W}[x] \nonumber \\
&+\alphacat \sum_{u,v}\pin \pout \ln \Bigl(\int Dx D\hat{x} P_0(x(0)) \nonumber \\
&\times \exp \qty[i\int dt A_{\rm cat}(x(t), \hat{x}(t), h_{{\rm cat},uv}(t), \psi_{{\rm cat},uv}(t)) - i\hat{U}_{{\rm cat},u} [\hat{x}] - i\hat{V}_{{\rm cat},v} [x, \hat{x}] - i\nu\hat{W} [x]]\Bigr) \nonumber \\
&+\alphanut \sum_{u,v}\pin \pout \ln \Bigl(\int Dx D\hat{x} P_0(x(0)) \nonumber \\
&\times\exp \qty[i\int dt A_{\rm nut}(x(t), \hat{x}(t), h_{{\rm nut},uv}(t), \psi_{{\rm nut},uv}(t)) - i\hat{U}_{{\rm nut},u} [\hat{x}] - i\hat{V}_{{\rm nut},v} [x, \hat{x}]]\Bigr).
\label{eq:SM_Phi}
\end{align}
Here, we assume that $h_i(t)$ and $\psi_i(t)$ depend only on the in-degree $u$, out-degree $v$, and group $\mathcal{G}_\sigma$ of the $i$-th chemical species; this assumption leads to the factorized form of the last term in Eq.~\eqref{eq:SM_Phi}.

The integral in Eq.~\eqref{eq:SM_GF3} can be evaluated by applying the saddle-point method in the large $N$ limit, and then the resulting saddle-point equations read
\begin{align}
\hat{U}_{\sigma, u}[\hat{x}] &= i \sum_{v}\pout \int Dy D\hat{y}Dz_1\cdots Dz_\nu (\sum_{\sigma} \alpha_\sigma V_{\sigma, v}[y, \hat{y}])W[z_1]\cdots W[z_\nu] f_{uv}[\hat{x}, y, \hat{y}, \{z\}] \equiv \hat{U}_{u}[\hat{x}], \label{eq:SM_saddle1} \\
\hat{V}_{\sigma,v}[y,\hat{y}] &= i \sum_{u}\pin \int D\hat{x} Dz_1 \cdots Dz_\nu (\sum_{\sigma} \alpha_\sigma U_{\sigma, u}[\hat{x}]) W[z_1]\cdots W[z_\nu] f_{uv}[\hat{x}, y, \hat{y}, \{z\}] \equiv \hat{V}_{v}[y,\hat{y}], \label{eq:SM_saddle2} \\
\hat{W}[z] &= i \sum_{u,v} \pin \pout \int D\hat{x} Dy D\hat{y}Dz_1\cdots Dz_{\nu-1} (\sum_{\sigma} \alpha_{\sigma} U_{\sigma,u}[\hat{x}]) (\sum_{\sigma} \alpha_\sigma V_{\sigma,v}[y, \hat{y}])W[z_1]\cdots W[z_{\nu-1}] f_{uv}[\hat{x}, y, \hat{y}, \{z\}], \label{eq:SM_saddle3} \\
U_{\sigma,u}[\hat{x}] &= \sum_{v} \pout \int Dx \gamma_\sigma[x,\hat{x}|u,v], \label{eq:SM_saddle4} \\
V_{\sigma,v}[y, \hat{y}] &= \sum_{u} \pout \gamma_\sigma[y,\hat{y}|u,v], \label{eq:SM_saddle5} \\
W[z] &= \sum_{u,v} \pin \pout \int D\hat{z} \gamma_{\rm cat}[z,\hat{z}|u,v], \label{eq:SM_saddle6}
\end{align}
where
\begin{align}
\gamma_{\rm cat}[x,\hat{x}|u,v] &= \frac{\exp \qty[i\int dt A_{\rm cat}(x(t), \hat{x}(t), h_{{\rm cat},uv}(t), \psi_{{\rm cat},uv}(t)) - i\hat{U}_{u}[\hat{x}] - i\hat{V}_{v}[x,\hat{x}] - i\nu\hat{W}[x]]}{\int Dx D\hat{x} \exp \qty[i\int dt A_{\rm cat}(x(t), \hat{x}(t), h_{{\rm cat},uv}(t), \psi_{{\rm cat},uv}(t)) - i\hat{U}_{u}[\hat{x}] - i\hat{V}_{v}[x,\hat{x}] - i\nu\hat{W}[x]]},
\label{eq:SM_gamma_cat} \\
\gamma_{\rm nut}[x,\hat{x}|u,v] &= \frac{\exp \qty[i\int dt A_{\rm nut}(x(t), \hat{x}(t), h_{{\rm nut},uv}(t), \psi_{{\rm nut},uv}(t)) - i\hat{U}_{u}[\hat{x}] - i\hat{V}_{v}[x,\hat{x}]]}{\int Dx D\hat{x} \exp \qty[i\int dt A_{\rm nut}(x(t), \hat{x}(t), h_{{\rm nut},uv}(t), \psi_{{\rm nut},uv}(t)) - i\hat{U}_{u}[\hat{x}] - i\hat{V}_{v}[x,\hat{x}]]}.
\label{eq:SM_gamma_nut}
\end{align}
It should be noted that the saddle-point equations for the conjugate (hatted) order parameters $\hat{U}_{\sigma, u}$ and $\hat{V}_{\sigma, v}$ do not depend on the species group $\sigma$; therefore, we introduce the shorthand notation, $\hat{U}_{u}[\hat{x}]$ and $\hat{V}_{v}[y, \hat{y}]$. 

To interpret the probability distribution $\gamma_\sigma[x,\hat{x}|u,v]$, we define the expectation with respect to $\gamma_\sigma[x,\hat{x}|u,v]$ as
\begin{align}
\expval{\cdots}_{\sigma, uv} = \int Dx D\hat{x} \cdots \gamma_\sigma[x,\hat{x}|u,v].
\end{align}
By calculating the functional derivatives of the expression for the generating functional $\ovl{\mathcal{Z}[\bm{h},\bm{\psi}]}$ in Eq.~\eqref{eq:SM_GF3} with respect to $\psi_{\sigma, uv}(t)$ and $h_{\sigma, uv}(t)$, we find the following relations:
\begin{align}
\expval{x(t)}_{\sigma, uv} = \int Dx D\hat{x} x(t) \gamma_\sigma [x,\hat{x}|u,v] 
&=  (-i) \lim_{\bm{h},\bm{\psi}\to 0} \frac{\delta \ovl{\mathcal{Z}[\bm{h},\bm{\psi}]}}{\delta \psi_{\sigma, uv}(t)} \nonumber \\
&=  (-i) \lim_{\bm{h},\bm{\psi}\to 0} \frac{1}{\alpha_\sigma \pin \pout N}\sum_{i \in \mathcal{G}_\sigma} \delta_{u_i,u} \delta_{v_i,v} \frac{\delta \ovl{\mathcal{Z}[\bm{h},\bm{\psi}]}}{\delta \psi_i(t)} \nonumber \\
&= \frac{1}{\alpha_\sigma \pin \pout N}\sum_{i \in \mathcal{G}_\sigma} \delta_{u_i,u} \delta_{v_i,v} \expval{x_i(t)}_\xi, \label{eq:SM_moment2} \\
\expval{\hat{x}(t)}_{\sigma, uv} = \int Dx D\hat{x} \hat{x}(t) \gamma_\sigma[x,\hat{x}|u,v] 
&=  i \lim_{\bm{h},\bm{\psi}\to 0} \frac{\delta \ovl{\mathcal{Z}[\bm{h},\bm{\psi}]}}{\delta h_{\sigma, uv}(t)} \nonumber \\
&=  i \lim_{\bm{h},\bm{\psi}\to 0} \frac{1}{\alpha_\sigma \pin \pout N}\sum_{i \in \mathcal{G}_\sigma} \delta_{u_i,u} \delta_{v_i,v} \frac{\delta \ovl{\mathcal{Z}[\bm{h},\bm{\psi}]}}{\delta h_i(t)} \nonumber \\
&= \frac{1}{\alpha_\sigma \pin \pout N}\sum_{i \in \mathcal{G}_\sigma} \delta_{u_i,u} \delta_{v_i,v} \expval{\hat{x}_i(t)}_\xi = 0, \label{eq:SM_moment_hat2} \\
\expval{x(t)\hat{x}(t')}_{\sigma, uv} = \int Dx D\hat{x} x(t) \hat{x}(t') \gamma_\sigma[x,\hat{x}|u,v] &= \frac{1}{\alpha_\sigma \pin \pout N}\sum_{i \in \mathcal{G}_\sigma} \delta_{u_i,u} \delta_{v_i,v} \expval{x(t)\hat{x}_i(t')}_\xi. \label{eq:SM_responce2}
\end{align}
Comparing these relations with Eqs.~\eqref{eq:SM_moment}-\eqref{eq:SM_response}, one can regard $\int D\hat{x} \gamma_\sigma[x,\hat{x}|u,v]$ as the path probability describing the typical dynamics of the abundance of the chemical species belonging to $\mathcal{G}_\sigma$, conditioned on the in-degree $u$ and out-degree $v$.

Using  Eqs.~\eqref{eq:SM_moment2}-\eqref{eq:SM_responce2} and substituting Eqs.~\eqref{eq:SM_fuvw2}, \eqref{eq:SM_saddle4}-\eqref{eq:SM_saddle6} into Eq.~\eqref{eq:SM_saddle3}, we find
\begin{align}
\hat{W}[z] &= \sum_{u,v}\pin \pout \frac{uv}{c^2} \Biggl[\int dt (\sum_{\sigma}\alpha_\sigma \sum_{v'} p_{v'}^{\rm (out)} \expval{\hat{x}(t)}_{\sigma, uv'})(\sum_{\sigma}\alpha_\sigma \sum_{v'} p_{u'}^{\rm (in)} \expval{y(t)}_{\sigma, u'v})\expval{z(t)}_{{\rm cat},uv}^{\nu-1} z(t) \nonumber \\
&~~~~~~~~~~~~~~~~~~~~~~~~~~~~~~~~~~~~~~~~~~~~~~~~~~~~~~~~~~~~~~~~~~~~~- (\sum_{\sigma} \alpha_\sigma \sum_{u'} p_{u'}^{\rm (in)} \expval{y(t)\hat{y}(t)}_{\sigma, u'v})\expval{z(t)}_{{\rm cat},uv}^{\nu-1} z(t)\Biggr] \nonumber \\
&= 0.
\label{eq:SM_w_hat}
\end{align}
Here, we use $\expval{\hat{x}(t)}_{\sigma,uv}=0$ and $\expval{\hat{x}(t)x(t)}_{\sigma, uv}=0$.
Vanishing of $\hat{W}[z]$ implies that the catalyzed chemical species do not provide feedback to the catalytic chemical species through the catalytic activity. Similarly to $\hat{W}[z]$, the other hatted order parameters can be calculated as
\begin{align}
\hat{U}_u[\hat{x}] &= \sum_{v}\pout \frac{uv}{c^2}\Biggl[\int dt \hat{x}(t)(\sum_{\sigma}\alpha_\sigma\sum_{u'} p_{u'}^{\rm (in)}\expval{y(t)}_{\sigma, u'v})(\sum_{u', v'} p_{u'}^{\rm (in)} p_{v'}^{\rm (out)}\expval{z(t)}_{{\rm cat},u'v'})^\nu \nonumber \\
&~~~~~~~~~~~~~~~~~~~~~~~~~~~~~~~~~~~~~~~~~~-(\sum_{\sigma}\alpha_\sigma\sum_{u'} p_{u'}^{\rm (in)}\expval{\hat{y}(t)y(t)}_{\sigma, u'v})(\sum_{u', v'} p_{u'}^{\rm (in)} p_{v'}^{\rm (out)}\expval{z(t)}_{{\rm cat},u'v'})^\nu \Biggr] \nonumber \\
&= \frac{u}{c} \int dt \hat{x}(t) m_{\rm in}(t) \mcat^\nu(t), \label{eq:SM_u_hat} \\
\hat{V}_v[y, \hat{y}] &= \sum_{u} \pin \frac{uv}{c^2}\Biggl[\int dt (\sum_{\sigma}\alpha_\sigma\sum_{v'} p_{v'}^{\rm (out)}\expval{\hat{x}(t)}_{\sigma, uv'})y(t)(\sum_{u',v'} p_{u'}^{\rm (in)} p_{v'}^{\rm (out)}\expval{z(t)}_{{\rm cat},u'v'})^\nu\nonumber \\
&~~~~~~~~~~~~~~~~~~~~~~~~~~~~~~~~~~~~~~~~~~~~~~~~~~~~~~~~~~~~~~~~~~~~~~~-\hat{y}(t)y(t)(\sum_{u',v'} p_{u'}^{\rm (in)} p_{v'}^{\rm (out)}\expval{z(t)}_{{\rm cat},u'v'})^\nu\Biggr] \nonumber \\
&=- \frac{v}{c} \int dt \hat{y}(t) y(t) \mcat^\nu(t), \label{eq:SM_v_hat}  
\end{align}
where
\begin{align}
m_{\rm in}(t) &= \sum_{\sigma, u, v} \frac{v}{c} \alpha_\sigma \pin \pout \expval{x(t)}_{\sigma, uv}, \label{eq:SM_min} \\
\mcat(t) &= \sum_{u,v} p_{u}^{\rm (in)} p_{v}^{\rm (out)}\expval{x(t)}_{{\rm cat},uv}. \label{eq:SM_mcat}
\end{align}
Substituting Eqs.~\eqref{eq:SM_u_hat} and \eqref{eq:SM_v_hat} into $\gamma_\sigma[x,\hat{x}|u,v]$, we obtain
\begin{align}
\gamma_\sigma[x,\hat{x}|u,v] = \frac{\exp \qty[i\int dt A_\sigma(x(t), \hat{x}(t), h_{\sigma, uv}(t), \psi_{\sigma, uv}(t))-i\hat{x}(t)(\frac{u}{c}m_{\rm in}(t) \mcat^\nu(t)-\frac{v}{c}x(t)\mcat^\nu(t))]}{\int DxD\hat{x}\exp \qty[i\int dt A_\sigma(x(t), \hat{x}(t), h_{\sigma, uv}(t), \psi_{\sigma, uv}(t))-i\hat{x}(t)(\frac{u}{c}m_{\rm in}(t) \mcat^\nu(t)-\frac{v}{c}x(t)\mcat^\nu(t))]}.
\end{align}
The path probability of the effective dynamical equation for the abundance of catalytic chemical species and nutrients, conditioned on the in-degree $u$ and out-degree $v$, can then be written as
\begin{align}
\nonumber
\mathcal{P}_{{\rm cat}, uv}[x, h] = &P_{{\rm cat}, uv0}(x(0))\int D\xi \mathcal{P}_\eta [\xi] \\
&\times\delta_{\rm F}\qty[\dot{x}(t)-\frac{u}{c} m_{\rm in}(t) \mcat^\nu (t) + \frac{v}{c} \mcat^\nu (t) x(t)+\mu(t)x(t) -h(t) - \xi(t)], \\
\nonumber
\mathcal{P}_{{\rm nut}, uv}[x, h] = &P_{{\rm nut}, uv0}(x(0))\int D\xi \mathcal{P}_\eta [\xi] \\
&\times\delta_{\rm F}\qty[\dot{x}(t)-\frac{u}{c} m_{\rm in}(t) \mcat^\nu (t) + \frac{v}{c} \mcat^\nu (t) x(t)- d(g - x(t))+\mu(t)x(t) -h(t) - \xi(t)],
\end{align}
respectively.
Hence, we finally obtain the effective dynamical equations as
\begin{align}
\dot{x}_{{\rm cat}, uv}(t) &= \frac{u}{c} m_{\rm in}(t) \mcat (t) - \frac{v}{c} \mcat (t) x_{{\rm cat}, uv}(t) - \mu(t)x_{{\rm cat}, uv}(t) + h_{{\rm cat}, uv}(t) + \xi_{{\rm cat}, uv}(t),  \label{eq:SM_dmfteq_cat} \\
\dot{x}_{{\rm nut}, uv}(t) &= \frac{u}{c} m_{\rm in}(t) \mcat (t) - \frac{v}{c} \mcat (t) x_{{\rm nut}, uv}(t) + d(g - x_{{\rm nut}, uv}(t)) - \mu(t)x_{{\rm nut}, uv}(t) + h_{{\rm nut}, uv}(t) + \xi_{{\rm nut}, uv}(t), \label{eq:SM_dmfteq_nut}
\end{align}
respectively. In the main text, we analyze the case of $h_{\sigma, uv}(t) = \xi_{\sigma, uv}(t) = 0$.


\section{Linear stability analysis}
\label{sec:stability}
To examine the stability of the overnutrition state for $\nu \ge 2$, which emerges when $\alphanut g > 1$,  we linearize the DMFT equations, given by Eq.~\eqref{eq:SM_dmfteq_cat} and Eq.~\eqref{eq:SM_dmfteq_nut}, around the fixed point by introducing a small perturbation $\xi_{\sigma, uv}(t)$~\cite{opper1992phase, roy2019numerical, galla2024generating}. In the case of $\nu=1$, which has been reported in Ref.~\cite{mitsumoto2026starvation}, the overnutrition state becomes unstable at $d<d_{\rm c}=1/(\alphanut g -1)$. The perturbation is assumed to be an independent white Gaussian noise for each $\sigma$, $u$, and $v$. We define the deviations from the fixed point as $\delta x_{\sigma, uv}(t) = x_{\sigma, uv}(t) - x_{\sigma, uv}^*$, $\delta \mcat(t) = \mcat(t) - \mcat^*$, $\delta m_{\rm in}(t) = m_{\rm in}(t) - m_{\rm in}^*$, and $\delta \mu(t) = \mu(t) - \mu^*$. If the solution is stable, all these deviations fluctuate around zero in the long-time limit.

The linearized dynamics around a given fixed point are written as
\begin{align}
\delta \dot{x}_{\sigma, uv}(t) &= \frac{u}{c}(\nu m_{\rm in}^*\delta \mcat(t) (\mcat^*)^{\nu-1} + \delta m_{\rm in}(t) (\mcat^*)^\nu) - \frac{v}{c}((\mcat^*)^\nu\delta x_{\sigma, uv}(t) + \nu\delta \mcat(t) (\mcat^*)^{\nu-1} x_{\sigma, uv}^*) \nonumber \\
&-(d_\sigma+\mu^*)\delta x_{\sigma, uv}(t)-\delta\mu(t) x_{\sigma, uv}^* + \xi_{\sigma, uv}.
\label{eq:linear}
\end{align}
In the overnutrition state, $\mcat^* = 0$; hence, Eq.~\eqref{eq:linear} becomes $\delta \dot{x}_{\sigma, uv}(t) = -(d_\sigma+\mu^*)\delta x_{\sigma, uv}(t)-\delta\mu(t) x_{\sigma, uv}^* + \xi_{\sigma, uv}$.
Denoting by $\tilde{f}(\omega)$ the Fourier transform of a function $f(t)$, Eq.~\eqref{eq:linear} can be rewritten as
\begin{align}
\delta \tilde{x}_{\sigma, uv}(\omega) = &\frac{\delta \tilde{\mu}(\omega)x_{\sigma, uv}^* - \tilde{\xi}_{\sigma, uv}(\omega)}{i \omega + d_\sigma + \mu^*}.
\label{eq:linear_fourier}
\end{align}
Substituting the non-metabolic solution into Eq.~\eqref{eq:linear_fourier}, we obtain the following expressions for
\begin{align}
\delta \tilde{x}_{{\rm cat}, uv}(\omega) &= -\frac{\tilde{\xi}_{{\rm cat}, uv}(\omega)}{i \omega + d(\alphanut g -1)}, \\
\delta \tilde{x}_{{\rm nut}, uv}(\omega) &= \frac{\delta \tilde{\mu}(\omega)/\alphanut - \tilde{\xi}_{{\rm nut}, uv}(\omega)}{i \omega + d\alphanut g}.
\end{align}
We then square these expressions and average over the Gaussian noise $\xi_{\sigma, uv}$ to obtain
\begin{align}
\expval{|\delta \tilde{x}_{{\rm cat}, uv}(\omega)|^2} &= \frac{I_{{\rm cat}, uv}}{\omega^2 + d^2(\alphanut g -1)^2}, \label{eq:delta_x1uv} \\
\expval{|\delta \tilde{x}_{{\rm cat}, uv}(\omega)|^2} &= \frac{\expval{|\delta \tilde{\mu}(\omega)/\alphanut|^2} + I_{{\rm nut}, uv}}{\omega^2 + d^2\alphanut^2 g^2}, \label{eq:delta_x2uv}
\end{align}
where $\expval{\cdots}$ denotes the average over $\xi_{\sigma, uv}$ and $I_{\sigma, uv} = \expval{|\tilde{\xi}_{\sigma, uv}|^2}\ge0$. In the limit $\omega \to 0$, one finds that $\expval{|\delta \tilde{x}_{{\rm cat},uv}(\omega)|^2}$ diverges at $\alphanut g = 1$, leading to the instability of overnutrition state for any permeability $d$ at $\alphanut g = 1$. From the definition of the growth rate $\mu$, we obtain
\begin{align}
\expval{|\delta \tilde{\mu} (\omega)|^2} = (d\alphacat)^2 \expval{|\delta \tilde{m}_{\rm cat}(\omega)|^2} \sim 1/(\omega^2+d^2(\alphanut g -1)^2),
 \end{align}
 which also diverges at $\alphanut g = 1$ in the limit $\omega \to 0$; hence, Eqs.~\eqref{eq:delta_x1uv} and \eqref{eq:delta_x2uv} diverge only when $\alphanut g = 1$. We therefore find that the overnutirition state is stable for any $d$ when $\alphanut g >1$.

\section{Asymptotic solutions for low nutrient supply}
\label{sec:asymptotic}

In this section we consider the low--growth situation discussed in Sec.~\ref{sec:starvation}, where $\mu^* \ll \alphanut g \ll 1$ and $\mu^* \ll d$. We show that, for both the uniform distribution with $r=1$ and the power-law distribution, 
the asymptotic behavior of $\mu^*$ as a function of $\alphanut g$ exhibits the exponential form $\exp(-C/\alphanut g)$, where $C$ is a positive constant.

The average abundance of catalytic chemical species and nutrients are given by $\mcat^* = m_{\rm in}^* (\mcat^*)^\nu X(\mu^*, (\mcat^*)^\nu)$ and $\mnut^* = (dg+m_{\rm in}^* (\mcat^*)^\nu)X(\mu^*+d, (\mcat^*)^\nu)$ in the main text, respectively, where $X(a,b) = \sum_{v}\pout/(a+\frac{v}{c}b)$. Hence, if $\mu^* \ll d$, fixed-point values of the average abundances of catalytic chemical species and nutrients are given by
\begin{align}
\mcat^* &= X(\mu^*,(\mcat^*)^\nu)m_{\rm in}^*(\mcat^*)^\nu, \label{eq:m1asympt} \\
\mnut^* & \simeq X(d,(\mcat^*)^\nu)(dg+m_{\rm in}^*(\mcat^*)^\nu), \label{eq:m2asympt} 
\end{align}
respectively. Using Eq.~\eqref{eq:growth1}, we also find
\begin{align}
\mcat^* \simeq \frac{1-\alphanut g}{\alphacat}.
\label{eq:mcatasympt}
\end{align}
Substituting this into Eq.~\eqref{eq:m1asympt}, we obtain the relation
\begin{align}
X(\mu^*,(\mcat^*)^\nu) \simeq \frac{1-\alphanut  g}{\alphacat m_{\rm in}^*(\mcat^*)^\nu}.
\label{eq:xmumcat}
\end{align}
Because the uniform distribution with $r=1$ and the power-law distribution with $\beta=3$ with dense connectivity, $c\gg1$, are written as $p(v/c)=1/2$ with the interval $[0,2]$ and $p(v/c)=2/(1+v/c)^3$ with the interval $[0,\infty)$, respectively, the function $X(a,b)$ can be evaluated as
\begin{align}
X(a,b) &= \sum_v \pout/(a+\frac{v}{c}b) \nonumber \\
&=
\begin{cases}
&\int_0^2 dv' 1/2(a+v'b)~~~~~~~~~~~~(\text{uniform}) \\
&\int_0^\infty 2/(1+v')^3(a+v'b)~~~~(\text{power-law})
\end{cases} \nonumber \\
&=
\begin{cases}
&\frac{1}{2b}\log\frac{a+2b}{a}~~~~~~~~~~~~~~~~~~~~~~(\text{uniform}) \\
&\frac{1}{(a-b)^2}\qty(\frac{2b^2}{a-b}\log\frac{b}{a}+a-3b)~(\text{power-law}),
\end{cases}
\end{align}
where $v'=v/c$. One finds that the value of $X(a,b)$ logarithmically diverges as $a\to+0$.
Thus, when $\mu^* \ll \alphanut g \ll 1$, the relation given by Eq.~\eqref{eq:xmumcat} leads to
\begin{align}
\mu^* =
\begin{cases}
&2(\mcat^*)^\nu \exp\qty(-2\frac{1-\alphanut g}{\alphacat m_{\rm in}^*})~~~~~(\text{uniform}) \\
&(\mcat^*)^\nu\exp\qty(-\frac{3}{2}-\frac{1-\alphanut g}{2\alphacat m_{\rm in}^*})~~(\text{power-law}).
\end{cases}
\label{eq:mu_relation}
\end{align}

Substituting Eqs. \eqref{eq:mcatasympt} and \eqref{eq:m2asympt} into Eq.~\eqref{eq:constraint2}, we can derive $m_{\rm in}^*$ as a function of $\alphanut g$ as
\begin{align}
m_{\rm in}^* = \frac{(1-dX(d,(\mcat^*)^\nu))\alphanut g}{(1-\alpha_1)(\mcat^*)^\nu X(d,(\mcat^*)^\nu)},
\label{eq:min1}
\end{align}
which is proportional to $\alphanut g$. Thus, the asymptotic behavior of $\mu^*$ as a function of $\alphanut g$ exhibits the exponential form $\exp(-C/\alphanut g)$.

\twocolumngrid

\bibliography{fk_model_dmft}

\end{document}